\documentclass[11pt]{article}
\usepackage[a4paper, margin=2cm]{geometry}
\usepackage[T1]{fontenc}
\usepackage[utf8]{inputenc}
\usepackage{authblk}
\usepackage{graphics}
\usepackage{graphicx}
\usepackage{subfig}
\usepackage{lscape}
\usepackage{tabulary}
\usepackage{booktabs}
\usepackage{longtable}
\usepackage{xfrac}
\usepackage{rotating}
\usepackage{fancybox}
\usepackage{alphalph}
\usepackage{parskip}
\usepackage{amsmath}
\usepackage{mathtools}
\usepackage{cite}

\providecommand{\keywords}[1]{\textbf{\textit{Keywords:}} #1}
\providecommand{\abbreviations}[1]{\textbf{\textit{Abbreviations:}} #1}
\title{Predictable patterns of CTL escape and reversion across host populations and viral subtypes in HIV-1 evolution}
\author[1,2,3]{Duncan S. Palmer\thanks{duncan.stuart.palmer@gmail.com}}
\author[4]{Emily Adland}
\author[2,5,6]{John Frater}
\author[4,7]{Philip J. R. Goulder}
\author[7]{Thumbi Ndung'u}
\author[5]{Philippa C. Matthews}
\author[2,5]{Rodney E. Phillips}
\author[8,9]{Roger Shapiro}
\author[1,3]{Gil McVean}
\author[2,10]{Angela R. McLean}
\affil[1]{Department of Statistics, 1 South Parks Road, University of Oxford, Oxford OX1 3TG, UK}
\affil[2]{Institute for Emerging Infections, The Oxford Martin School, Oxford OX1 3BD, UK}
\affil[3]{Wellcome Trust Centre for Human Genetics, Roosevelt Drive, Oxford OX3 7BN, UK}
\affil[4]{Department of Paediatrics, University of Oxford, Oxford OX1 3SY, UK}
\affil[5]{Nuffield Department of Clinical Medicine, Peter Medawar Building for Pathogen Research, University of Oxford, Oxford OX1 3SY, UK}
\affil[6]{Oxford NIHR Biomedical Research Centre, Oxford OX3 7LE, UK}
\affil[7]{HIV Pathogenesis Programme, Doris Duke Medical Research Institute, University of KwaZulu-Natal, Durban, 4013 South Africa}
\affil[8]{Botswana Harvard AIDS Institute Partnership, Gaborone BO 320, Botswana}
\affil[9]{Department of Immunology and Infectious Diseases, Harvard School of Public Health, Boston, MA 02215}
\affil[10]{Zoology Department, South Parks Road, University of Oxford, Oxford OX1 3PS, UK}

\date{\vspace{-5ex}}
\begin{document}
\graphicspath{{../generated_figures/}}
\newcommand{\geogHOMERSSITTescp}{0.00000675}
\newcommand{\geogHOMERSSITTescrho}{0.825}
\newcommand{\geognegremHOMERSSITTescp}{0.000189}
\newcommand{\geognegremHOMERSSITTescrho}{0.779}
\newcommand{\geogHOMERSSITTrevp}{0.0801}
\newcommand{\geogHOMERSSITTrevrho}{0.336}
\newcommand{\geogextremeremHOMERSSITTrevp}{0.00646}
\newcommand{\geogextremeremHOMERSSITTrevrho}{0.681}
\newcommand{\geogHOMERbloemescp}{0.0169}
\newcommand{\geogHOMERbloemescrho}{0.538}
\newcommand{\geognegremHOMERbloemescp}{0.0276}
\newcommand{\geognegremHOMERbloemescrho}{0.549}
\newcommand{\geogHOMERbloemrevp}{0.00129}
\newcommand{\geogHOMERbloemrevrho}{0.715}
\newcommand{\geogextremeremHOMERbloemrevp}{0.000193}
\newcommand{\geogextremeremHOMERbloemrevrho}{0.811}
\newcommand{\geogSSITTbloemescp}{0.00570}
\newcommand{\geogSSITTbloemescrho}{0.614}
\newcommand{\geognegremSSITTbloemescp}{0.00730}
\newcommand{\geognegremSSITTbloemescrho}{0.657}
\newcommand{\geogSSITTbloemrevp}{0.211}
\newcommand{\geogSSITTbloemrevrho}{0.216}
\newcommand{\geogextremeremSSITTbloemrevp}{0.0891}
\newcommand{\geogextremeremSSITTbloemrevrho}{0.467}
\newcommand{\geogSSITTafricaescp}{0.000175}
\newcommand{\geogSSITTafricaescrho}{0.765}
\newcommand{\geognegremSSITTafricaescp}{0.000170}
\newcommand{\geognegremSSITTafricaescrho}{0.819}
\newcommand{\geogSSITTafricarevp}{0.331}
\newcommand{\geogSSITTafricarevrho}{0.114}
\newcommand{\geogextremeremSSITTafricarevp}{0.329}
\newcommand{\geogextremeremSSITTafricarevrho}{0.257}
\newcommand{\geogHOMERafricaescp}{0.000963}
\newcommand{\geogHOMERafricaescrho}{0.711}
\newcommand{\geognegremHOMERafricaescp}{0.00240}
\newcommand{\geognegremHOMERafricaescrho}{0.723}
\newcommand{\geogHOMERafricarevp}{0.0298}
\newcommand{\geogHOMERafricarevrho}{0.466}
\newcommand{\geogextremeremHOMERafricarevp}{0.0216}
\newcommand{\geogextremeremHOMERafricarevrho}{0.700}
\newcommand{\geogbloemafricaescp}{0.00000112}
\newcommand{\geogbloemafricaescrho}{0.861}
\newcommand{\geognegrembloemafricaescp}{0.0000327}
\newcommand{\geognegrembloemafricaescrho}{0.831}
\newcommand{\geogbloemafricarevp}{0.0768}
\newcommand{\geogbloemafricarevrho}{0.341}
\newcommand{\geogextremerembloemafricarevp}{0.00538}
\newcommand{\geogextremerembloemafricarevrho}{0.817}

\newcommand{\fitnessSSITTrevextremesrho}{0.810}
\newcommand{\fitnessSSITTrevextremesp}{0.000225}
\newcommand{\fitnessbloemrevextremesrho}{0.411}
\newcommand{\fitnessbloemrevextremesp}{0.0507}
\newcommand{\fitnessSSITTescp}{0.0231}
\newcommand{\fitnessSSITTrevp}{0.458}
\newcommand{\fitnessHOMERescp}{0.0541}
\newcommand{\fitnessHOMERrevp}{0.00815}
\newcommand{\fitnessbloemescp}{0.467}
\newcommand{\fitnessbloemrevp}{0.256}
\newcommand{\fitnessafricaescp}{0.443}
\newcommand{\fitnessafricarevp}{0.0371}
\newcommand{\fitnessSSITTescrho}{-0.451}
\newcommand{\fitnessSSITTrevrho}{0.0252}
\newcommand{\fitnessHOMERescrho}{-0.370}
\newcommand{\fitnessHOMERrevrho}{0.530}
\newcommand{\fitnessbloemescrho}{-0.0196}
\newcommand{\fitnessbloemrevrho}{0.156}
\newcommand{\fitnessafricaescrho}{-0.0331}
\newcommand{\fitnessafricarevrho}{0.398}

\newcommand{\SSITTODEMAPescp}{0.00000102}
\newcommand{\SSITTODEMAPrevp}{0.0909}
\newcommand{\HOMERODEMAPescp}{0.0000245}
\newcommand{\HOMERODEMAPrevp}{0.00670}
\newcommand{\bloemODEMAPescp}{0.0000993}
\newcommand{\bloemODEMAPrevp}{0.0147}
\newcommand{\SSITTODEMAPescrho}{0.862}
\newcommand{\SSITTODEMAPrevrho}{0.320}
\newcommand{\HOMERODEMAPescrho}{0.794}
\newcommand{\HOMERODEMAPrevrho}{0.556}
\newcommand{\bloemODEMAPescrho}{0.753}
\newcommand{\bloemODEMAPrevrho}{0.500}

\newcommand{\fitnessdiffmethodsp}{0.172}
\newcommand{\fitnessdiffmethodsrho}{0.295}
\newcommand{\fitnessdiffmethodsppear}{0.0178}
\newcommand{\fitnessdiffmethodsr}{0.489}

\newcommand{\pescfitmulti}{0.00343}
\newcommand{\prevfitmulti}{0.0113}
\newcommand{\pescgeogmulti}{0}
\newcommand{\prevgeogmulti}{0.0011}
\newcommand{\pescgeognegmulti}{0.0006}
\newcommand{\prevgeognegmulti}{0.00772}

\maketitle
\begin{abstract}
\noindent The twin processes of viral evolutionary escape and reversion in response to host immune pressure, in particular the cytotoxic T-lymphocyte (CTL) response, shape Human Immunodeficiency Virus-1 sequence evolution in infected host populations. The tempo of CTL escape and reversion is known to differ between CTL escape variants in a given host population. Here, we ask: are rates of escape and reversion comparable across infected host populations? For three cohorts taken from three continents, we estimate escape and reversion rates at 23 escape sites in optimally defined Gag epitopes. We find consistent escape rate estimates across the examined cohorts. Reversion rates are also consistent between a Canadian and South African infected host population. Certain Gag escape variants that incur a large replicative fitness cost are known to revert rapidly upon transmission. However, the relationship between escape/reversion rates and viral replicative capacity across a large number of epitopes has not been interrogated. We investigate this relationship by examining {\it in vitro} replicative capacities of viral sequences with minimal variation: point escape mutants induced in a lab strain. Remarkably, despite the complexities of epistatic effects exemplified by pathways to escape in famous epitopes, and the diversity of both hosts and viruses, CTL escape mutants which escape rapidly tend to be those with the highest replicative capacity when applied as a single point mutation. Similarly, mutants inducing the greatest costs to viral replicative capacity tend to revert more quickly. These data suggest that escape rates in {\it gag} are consistent across host populations, and that in general these rates are dominated by site specific effects upon viral replicative capacity. 
\end{abstract}

\abbreviations{HIV-1, Human Immunodeficiency Virus-1; CTL, Cytotoxic T-lymphocyte, MCMC; Markov Chain Monte-Carlo}

\noindent In HIV-1 infected individuals there is a constant arms race between the virus on one side and the host immune response on the other \cite{citeulike:10114103,citeulike:6485609,citeulike:13527420}. The cytotoxic T-lymphocyte (CTL) response - a part of the adaptive immune system - is a major driving force of HIV evolution both within and across infected individuals \cite{citeulike:5342331,citeulike:13799763}. Within nucleated cells, cytosolic peptides (both self and non-self) are presented at the cell surface by human leukocyte antigen (HLA) class I molecules \cite{citeulike:13399916}, encoded by HLA genes; the most variable loci in the entire human genome \cite{citeulike:11544589}. These protein fragments are known as `epitopes' upon presentation. CTLs may then recognise epitopes as non-self and destroy the presenting cell. The CTL response can select for mutations within or flanking epitopes which result in reduced immune recognition of virally infected cells. These variants, termed CTL escape mutants, may reduce HLA class I binding affinity, alter epitope processing, or affect T-cell receptor contact sites \cite{citeulike:13399940,citeulike:9632653,citeulike:13399944,citeulike:11286998}. Importantly, the repertoire of viral epitopes that may be presented is dependent upon the `type' of the HLA class I molecule. If an epitope can be presented, it is said to be `restricted' by that HLA type, and the host is known as `HLA matched' for that epitope. A host lacking the class I molecule required to present the epitope is known as `HLA mismatched' for that epitope. Viruses bearing escape mutations can be transmitted between hosts \cite{citeulike:7807211,citeulike:9633446}, and reversion of the infecting virus may take place due to the removal of HLA dependent selection pressure and viral fitness costs of CTL escape mutants \cite{citeulike:9405209,citeulike:7807208,citeulike:5342132,citeulike:13801183}.

The viral population present in a given infected individual reflects the selective environment of the current host as well as the remnants of selection from previous hosts. During the course of HIV-1 infection, viral mutations can open pathways to CTL immune escape or compensate for costs to replicative capacity of such mutations \cite{citeulike:5342137,citeulike:13597563,citeulike:3176628,citeulike:7807234,citeulike:6544540}, which can lead to a delay in reversion upon subsequent transmission \cite{citeulike:3176628}. Methods used to model and parameterise the CTL escape and subsequent reversion often assume that the timings of escape and reversion are governed by behaviour at a single site \cite{citeulike:9355716,citeulike:12345118,citeulike:9567453,citeulike:9956862,citeulike:13605117,citeulike:1540906,citeulike:4839667,citeulike:9778870}. This is a clear simplification. The existence and potential complexity of epistatic interactions, coupled with the diversity in HLA profiles, CTL killing efficiency, drug regimes, and vertical T-cell immunodominance \cite{citeulike:11847211} leads to the following questions: can individual escape mutants be characterised as having their own escape and reversion rates in a given infected host population, and are escape and reversion rates consistent across host populations?

The ideal data to answer these questions would consist of longitudinal viral sequence samples starting early in infection from a large number of hosts with known HLA types. However, this is not feasible. Instead, we have developed a method which allows us to use cross-sectional viral sequence data and host HLA information to estimate escape and reversion rates whilst accounting for the underlying viral genealogy \cite{citeulike:12345118}. This allows us to estimate population specific rates of escape and reversion at escape sites within optimally defined \cite{Brandereps} epitopes. By comparing the resultant escape/reversion rate estimates, we can test their consistency between infected host populations.

We apply our model to {\it gag} sequence data taken from three cohorts sampled from three continents: HOMER (British Columbia, Canada) \cite{citeulike:13587978}, SSITT (Switzerland) \cite{citeulike:9530950, citeulike:9405209}, and Bloemfontein (Bloemfontein, South Africa) \cite{citeulike:12828028,citeulike:9628954}. We find agreement between escape rate estimates, with the strongest positive correlation observed between CTL immune escape estimates across cohorts. Escape rate estimates are consistent between viral populations with distinct HIV-1 subtypes, where {\it{gag}} sequence divergence is as high as $\sim 9.2\%$. We also find consistent reversion rate estimates for the Bloemfontein and HOMER datasets (distinct HIV-1 subtypes, average nucleotide sequence divergence: $8.9\%$). Given the potential collection of paths through sequence space which could restore replicative fitness for a given escape variant, we find it particularly surprising to observe such a highly significant association between reversion rate estimates. These results suggest that given information regarding the dynamics of escape and reversion in one region, we may begin to make valid statements about these rate processes in other parts of the world, even when viral populations and HLA frequencies are substantially different. Our results support the findings in \cite{citeulike:7918948}, in which escape pathways are found to be broadly similar even across viral subtypes, and align with fitness effects in Pol observed by Hinkley {\it et al.} \cite{citeulike:9219371} which demonstrate that whilst HIV-1 protease and reverse transcriptase is `characterised by strong epistasis', a large portion of predictive power of replicative capacity is through single locus effects. To determine the extent of single locus effects on viral escape/reversion rates, we turn to data gathered from {\it in vitro} assessments of viral replicative capacity \cite{citeulike:12198074}. These estimates use site directed mutagenesis to induce escape variants in an otherwise conserved viral background, so do not reflect epistatic interactions in pathways to escape. We ask: do we observe correlations between our population derived escape or reversion rates, and these {\it in vitro} replicative capacity estimates? The emergence of a new consensus viral sequence within a given host is determined by the balance between immune escape and replicative fitness cost within that host's immunological background \cite{citeulike:1943760,citeulike:13799881,citeulike:10114103,citeulike:13799882, citeulike:13801178}. Naively ignoring all epistatic interactions, viral diversity, and assuming that all CTL escapes are equally beneficial to the virus, we would expect an escape mutation which incurs a small cost to viral replicative capacity to rapidly fix in the host's viral population. Similarly, an escape mutant which dramatically reduces replicative capacity would be expected to take far longer to reach intra-host consensus (we may make similar intuitive statements regarding reversion). Remarkably, despite these vast simplifying assumptions informing this intuition, we find a significant positive correlation between our SSITT escape rate estimates and {\it in vitro} replicative capacity, and a significant negative correlation between our HOMER reversion rate estimates and {\it in vitro} replicative capacity. We find it surprising to see such correlations between rate estimates measured in a reductionist, {\it in vitro} replicative capacity assay and a real world population of diverse individuals sharing diverse viruses. We conclude that whatever epistatic effects may be present across the HIV-1 proteome, in general rates of escape and reversion are largely dictated by the costs and benefits of individual mutations.
\section*{Methods}\label{section:chapter_4_methods}

\subsection*{Cohorts}
We consider paired viral sequence and host HLA type data taken from studies carried out in three populations: Switzerland (Swiss portion of the Swiss-Spanish intermittent treatment trial (SSITT) cohort \cite{citeulike:9530950, citeulike:9405209}); British Columbia, Canada (HAART observational medical evaluation and research (HOMER) cohort \cite{citeulike:13587978}); and Bloemfontein, South Africa \cite{citeulike:12828028,citeulike:9628954}. For each dataset bulk sequencing data was available, in which sequences obtained are assumed to represent the intra-host viral consensus sequence. We briefly outline the studies and portions of datasets used in Table \ref{table:cohort_summary}. For each dataset, we restrict our attention to {\it gag} and to the subset of sequences with the majority subtype in each population as assessed using RIP \cite{citeulike:12823836}: subtype B for SSITT and HOMER, and subtype C for Bloemfontein. Our reasons are threefold. Firstly, non-synonymous mutations in {\it gag} are likely to have a detrimental fitness effect as they encode structural proteins that are among the most highly conserved in the viral proteome \cite{citeulike:7916400}. Indeed, variants in {\it gag} can result in a measurable reduction in viral replicative capacity, as estimated by a variety of methods \cite{citeulike:13798492,citeulike:11611481,citeulike:13798494,citeulike:12198074,citeulike:13799912}. Secondly, the most protective HLA alleles are associated with CTL responses to Gag proteins \cite{citeulike:1027726, citeulike:13239698, citeulike:13239697}. Finally, site directed mutagenesis followed by replicative capacity assays have been carried out for a large number of known escape variants located in the {\it gag} gene \cite{citeulike:12198074}.

\begin{table}
\scriptsize
\begin{tabulary}{\textwidth}{l l l l}
  \toprule
  {\bf Dataset} & SSITT \cite{citeulike:9530950, citeulike:9405209} & HOMER \cite{citeulike:13587984} & Bloemfontein \cite{citeulike:12828028,citeulike:9628954} \\
  {\bf Population analysed} & Switzerland & British Columbia, Canada & Bloemfontein, South Africa\\
  {\bf Cohort size} & \vtop{\hbox{\strut $n=133$}\hbox{\strut Z{\"u}rich ($n=29$), Geneva ($n=26$),}\hbox{\strut Bern ($n=11$), Basel ($n=11$)}\hbox{\strut St Gall ($n=8$), Lugano ($n=7$)}\hbox{\strut Lausanne ($n=5$)}}& $n=567$ & \vtop{\hbox{\strut $n=884$,}\hbox{\strut plasma taken from $n=278$}}\\
  {\bf Sampling date} & 2000 & Between 1996 and 1999& February - September 2006 \\
  {\bf Sequences analysed} & p17 $n=38$, p24 $n=55$& $n=184$ & $n=198$\\
  {\bf Treatment} & HAART & \vtop{\hbox{\strut ART naive on recruitment,}\hbox{\strut initiated HAART between}\hbox{\strut August 1996 and September 1999}} & ART naive\\
  {\bf Study requirements} & \vtop{\hbox{\strut Undetectable VL for >6 months,}\hbox{\strut CD4$^+$ count $>300\mu l^{-1}$.}\hbox{\strut no history of non-nucleoside}\hbox{\strut reverse transcriptase inhibitors}} & $\geq 3$ antiretroviral drugs& \vtop{\hbox{\strut Chronic, plasma taken by}\hbox{\strut CD4$^+$ count,}\hbox{\strut low and high favoured:}\hbox{\strut 96 high ($>500\mu l^{-1}$),}\hbox{\strut 18 medium ($200-400\mu l^{-1}$),}\hbox{\strut 164 low ($<100\mu l^{-1}$)}}\\
\bottomrule
\end{tabulary}
\caption[Summary of three studies from which {\it gag} sequence data and HLA information was available]{Summary of three studies from which {\it gag} sequence data was available. For the SSITT and Bloemfontein datasets the number of sequences analysed is lower than the cohort size through a combination of lack of sequencing of {\it gag}, and data cleaning. For the HOMER dataset, the number of sequences analysed is lower than the cohort size due to restriction of the analysis to a single (unknown) calendar year, and data cleaning.\label{table:cohort_summary}}
\end{table}

Due to issues of confidentiality, we were unable to obtain the year of sequencing for the analysed portion of the HOMER dataset \cite{citeulike:13587978}. However, this is not required as we do not require a rescaling to calendar time (measured in {\it years}) from {\it substitutions site}$^{-1}$ to determine the existence of rank correlations of rate estimates across datasets. The 184 viral sequences which we analyse from HOMER dataset are taken from the year between 1996 and 1999 inclusive in which the largest number of {\it gag} sequences were obtained. For the SSITT dataset, separate collections of sequences were available for sequences encoding the p17 and p24 proteins in the {\it gag} gene. There was not a strong overlap between the patient identifiers of these two sequence sets, so we chose to analyse the two viral sequence regions independently.

\subsection*{Estimating escape and reversion rates}
We estimate HLA prevalence using data from the HLA FactsBook \cite{citeulike:7931760}. For the Canadian (HOMER) and Swiss (SSITT) datasets, we set the prevalence at the `Caucasian' estimate. For the African data (Bloemfontein, and the collection of cohorts summarised in Table \ref{table:african_cohort_summary}), we set the prevalence at the `Black' estimate. Epitopes examined by Boutwell {\it et al.} represent all optimally defined CTL epitopes at the time of publication \cite{Brandereps}. Escape mutations within these epitopes were defined based upon a list of HLA-associated HIV polymorphisms identified via statistical association \cite{citeulike:6544540,citeulike:12828039}. We define escape as any amino acid changes which occur at the same position as escape mutations provided in Boutwell {\it et al.} \cite{citeulike:12198074}. This is clearly an imperfect definition, but provides a sensible compromise between the clear overestimation of allowing all mutations within an epitope to be defined as escape, and only considering exact mutations validated as escapes {\it in vitro} \cite{citeulike:9355716}. In both of the approaches that we describe, it is important to note that estimates of escape represent averages across hosts known to restrict the epitope, and do not condition on the host making a CTL response or driving an escape during the course of infection. We are thus implicitly incorporating a scaling - the probability that an epitope is targetted conditional on the individual being HLA matched, which may vary across epitopes.

\subsubsection*{Maximum {\it a posteriori} (MAP) estimate using our integrated method}
Full details of the method are provided in Palmer {\it et al.} \cite{citeulike:12345118}. Briefly, we combine Felsenstein's tree peeling algorithm with existing phylogenetic software to merge statistical phylogenetic techniques with an ordinary differential equation (ODE) model which captures the processes of transmission, escape, and reversion. By doing so, we are able to use the information contained within the viral genealogy to infer estimates of escape and reversion rates ($\lambda_{\mathrm{esc}},\lambda_{\mathrm{rev}}$). There are four key steps in our inference regime:
\begin{enumerate}
\item Make the mild assumption that the genealogy and HLA/escape information are conditionally independent given the viral sequence information with the epitope removed.
\item We perform Markov chain Monte Carlo (MCMC) to sample genealogies from the posterior conditional on the viral sequence data with the epitope removed using BEAST \cite{citeulike:10387431}. We use a coalescent prior in an exponentially growing infected host population. The growth rate parameter of the infected population is sampled within the MCMC.
\item For each sampled viral genealogy, we determine the posterior density for $(\lambda_{\mathrm{esc}},\lambda_{\mathrm{rev}})$. This is achieved through a modification of Felsenstein's peeling algorithm \cite{citeulike:7890239}. Further details of this step can be found in the Supplementary Methods.
\item By averaging over tree specific posteriors, we obtain a consensus posterior density for $(\lambda_{\mathrm{esc}},\lambda_{\mathrm{rev}})$.
\end{enumerate}
\subsubsection*{Purely ODE based, naive model}
To analyse a particularly large fourth dataset for which rate estimation under our integrated model is not computationally feasible, we turn to a simpler compartment based ordinary differential equation (ODE) model of Fryer {\it et al.}. For each escape variant under consideration, viral sequence/HLA genotype pairs are split into four classes: HLA mismatched without escape, HLA mismatched with escape, HLA matched without escape and HLA matched with escape. Given these four proportions for an escape site, we then determine a best fit for $\lambda_{\mathrm{esc}}$ and $\lambda_{\mathrm{rev}}$ under the model described in Fryer {\it et al.} \cite{citeulike:9355716} during the exponential growth phase of the epidemic (Equations \eqref{equation_1} - \eqref{equation_4} in the Supplementary methods). Best estimates are determined using BFGS \cite{citeulike:8528122}, an approximation to Newton's method of hill climbing. Time of the start of the epidemic is set at the time to the most recent common ancestor of a BEAST run on the {\it gag} sequence data for each dataset.

Estimates of time to escape/reversion could not be obtained when no individuals in the cohort possessed the restricting HLA. When an escape mutation residing in an epitope reaches consensus in the population, the variant peptide is known as a negatope \cite{citeulike:9633474,citeulike:13587984}. Negatopes of HIV-1 subtype B as reported in Boutwell {\it et al.} \cite{citeulike:12198074} are highlighted in Tables \ref{SSITT_big_table} - \ref{bloemfontein_big_table} by asterisks.

\subsection*{{\it In vitro} viral replicative capacity}
In Boutwell {\it et al.} \cite{citeulike:12198074}, the fitness cost of an escape variant is approximated by determining the {\it in vitro} replicative capacity of the escape mutant inserted into a subtype B lab strain, in the absence of any other sequence variation. The measure of fitness is essentially the exponential {\it in vitro} growth rate of an isolate relative to the lab strain. Full details of the procedure are provided in \cite{citeulike:12198074}. We use these published replicative capacity assay measurements for 23 {\it gag} escape variants where both viral sequence data and the relevant HLA locus information was available in all three cohorts.
\subsection*{Accounting for multiple testing}
To account for multiple testing for each collection of tests of association, we perform permutation tests by shuffling the labellings of escape variants and re-evalaute correlation coefficients 100,000 times. This allows us to estimate the probability of observing a collection of Spearman rank correlation coefficients at least as extreme as those observed in the data. In all cases we obtain significant $p$-values. In each section, we also report the pairwise correlation coefficients, and describe a correlation as significant if its associated $p$-value is below 0.05. 
\section*{Results}\label{section:chapter_4_results}
We apply our integrated model \cite{citeulike:12345118}, and purely ODE based model \cite{citeulike:9355716} to data from the SSITT, HOMER and Bloemfontein cohorts. Detailed results, together with estimated replicative capacity of escape variants as previously estimated by an {\it in vitro} assay \cite{citeulike:12198074} are shown in Tables \ref{SSITT_big_table}, \ref{HOMER_big_table} and \ref{bloemfontein_big_table} for the HOMER, SSITT and Bloemfontein datasets respectively. Figure \ref{figure:HLA_proportions} summarises the prevalence of escape in all three regions. For each escape site we determine the proportion of HLA matched hosts harbouring the viral escape form. We perform the same calculation for HLA mismatched hosts. These two values for each of the 23 escape sites are then plotted against each other. The majority of the points lie below $y=x$, suggesting that selection for escape is taking place in the majority of the HLA/epitope pairings.

Estimates of time to escape and time to reversion for the HOMER dataset are given in units of {\it substitutions site}$^{-1}$. This is because the date of sampling was not available due to issues of confidentiality, and thus a scaling from {\it substitutions site}$^{-1}$ to {\it years} could not be estimated using BEAST \cite{citeulike:1888819}. Throughout our results, we report the Spearman rank correlation coefficient. We choose this statistic as variance in rate estimates increases drastically as the true underlying rates tend to infinity (or zero if we use a log scale). Also, when comparing replicative capacity to rate of escape or reversion, it is unclear {\it a priori} that any such relationship would be linear. In each plot, epitopes are abbreviated by writing the first amino acid of the epitope, followed by last, followed by the epitope length (e.g KRWIILGLNK $\rightarrow$ KK10).
\subsection*{Are rates of escape and reversion consistent across infected populations in different parts of the world?}

\begin{figure*}
\centering

\begin{minipage}[t]{0.05\textwidth}
\begin{flushleft}
\end{flushleft}
\end{minipage}
\begin{minipage}[t]{0.292\textwidth}
\begin{flushleft}
\quad\quad\quad\large{A}
\end{flushleft}
\end{minipage}
\begin{minipage}[t]{0.292\textwidth}
\begin{flushleft}
\quad\quad\quad\large{B}
\end{flushleft}
\end{minipage}
\begin{minipage}[t]{0.292\textwidth}
\begin{flushleft}
\quad\quad\quad\large{C}
\end{flushleft}
\end{minipage}

\vspace{-8mm}
\subfloat{\includegraphics[width=0.085\textwidth]{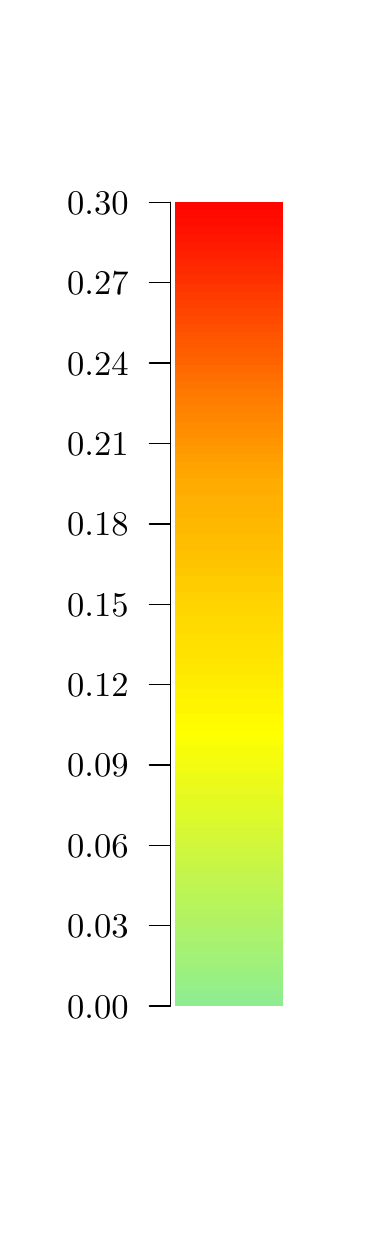}}\hspace{-0.025\textwidth}
\setcounter{subfigure}{0}
\subfloat{\label{figure:SSITT_HOMER}\includegraphics[width=0.3\textwidth]{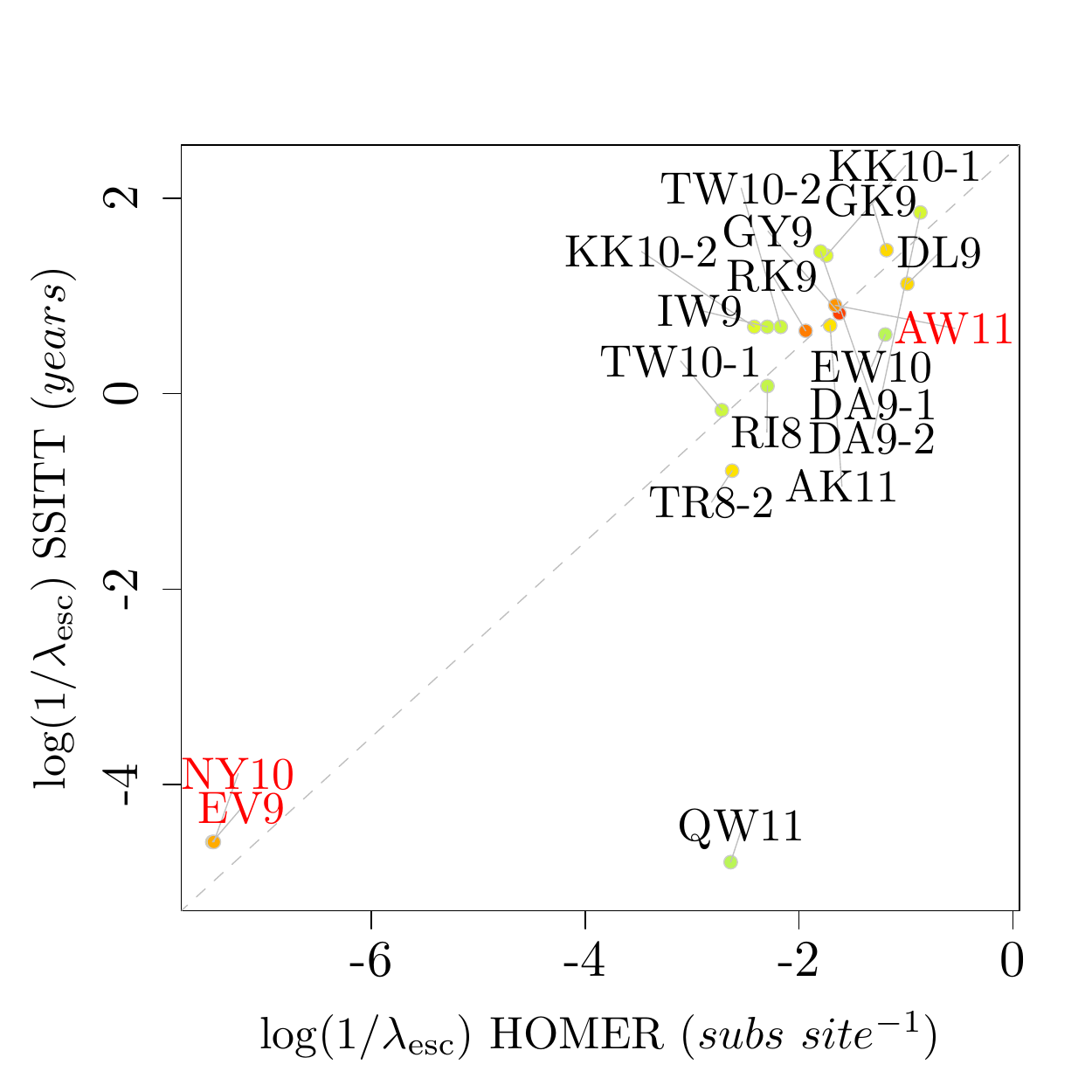}}
\subfloat{\label{fig:3a_geog}\includegraphics[width=0.3\textwidth]{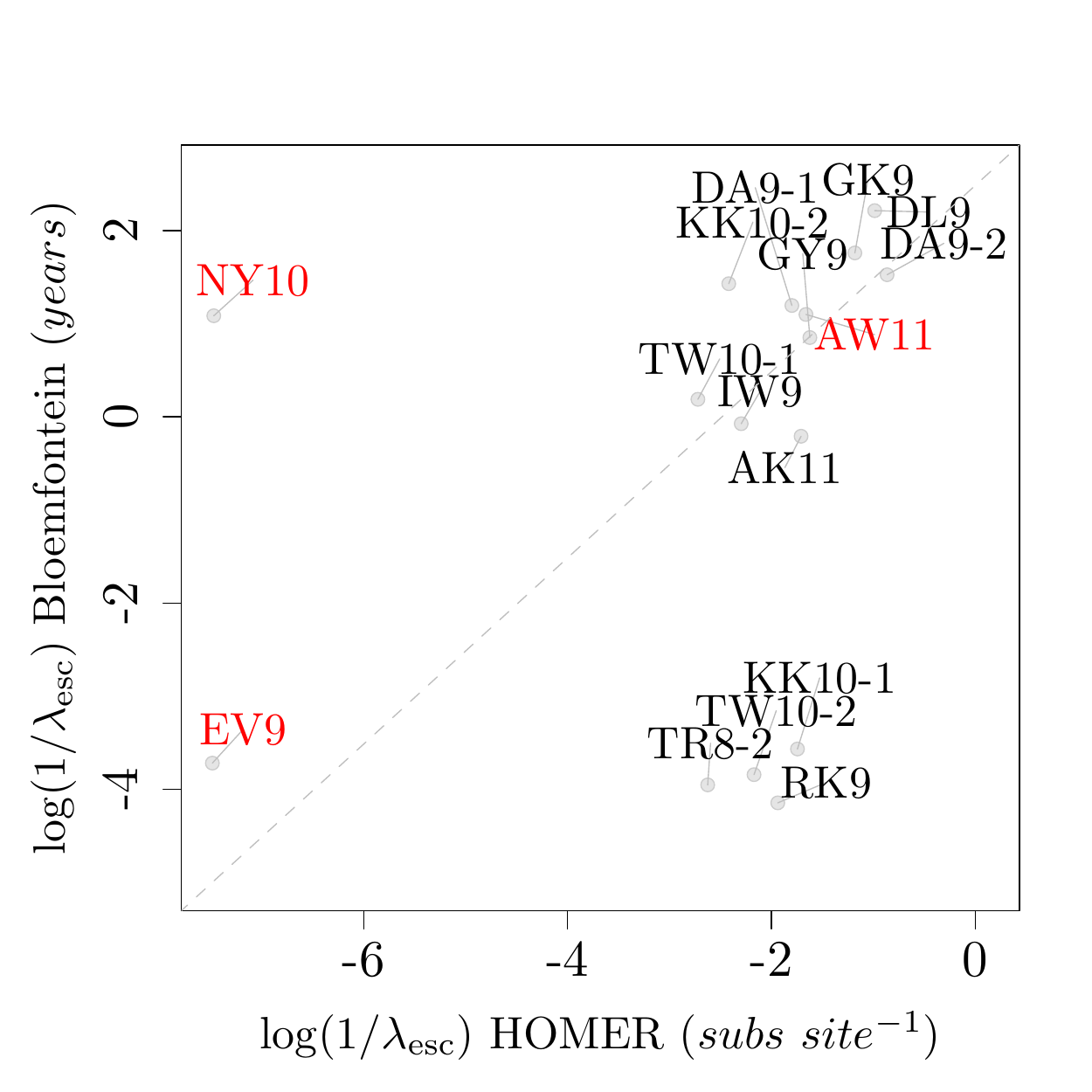}}
\subfloat{\label{fig:4a_geog}\includegraphics[width=0.3\textwidth]{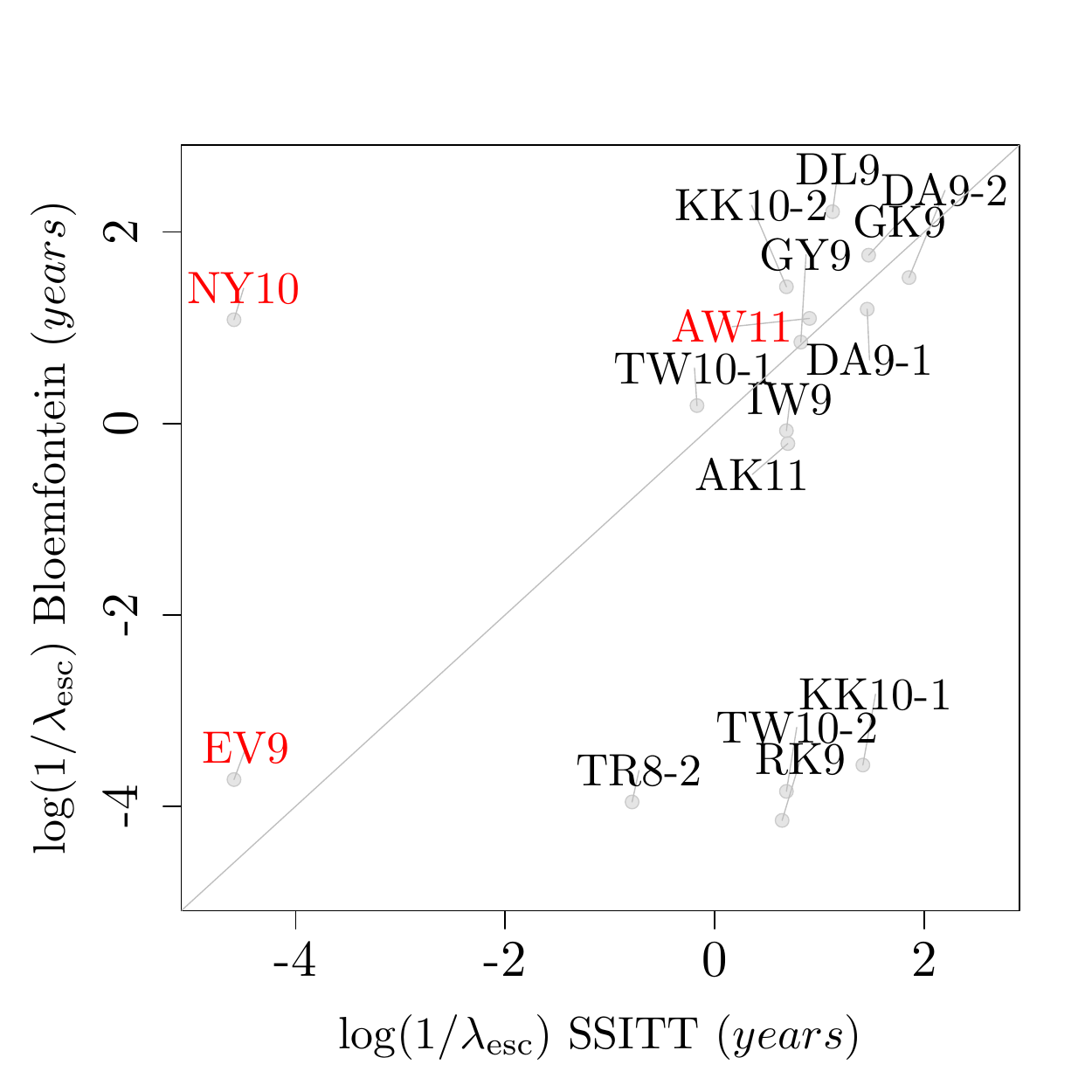}}\\\vspace{-2mm}
\subfloat{\makebox[0.06\textwidth]{}}
\subfloat{\makebox[0.3\textwidth]{$\rho=\geogHOMERSSITTescrho$; $p=\geogHOMERSSITTescp$.}}
\subfloat{\makebox[0.3\textwidth]{$\rho=\geogHOMERbloemescrho$; $p=\geogHOMERbloemescp$.}}
\subfloat{\makebox[0.3\textwidth]{$\rho=\geogSSITTbloemescrho$; $p=\geogSSITTbloemescp$.}}\\
\vspace{5mm}
\begin{minipage}[t]{0.05\textwidth}
\begin{flushleft}
\end{flushleft}
\end{minipage}
\begin{minipage}[t]{0.292\textwidth}
\begin{flushleft}
\quad\quad\quad\large{D}
\end{flushleft}
\end{minipage}
\begin{minipage}[t]{0.292\textwidth}
\begin{flushleft}
\quad\quad\quad\large{E}
\end{flushleft}
\end{minipage}
\begin{minipage}[t]{0.292\textwidth}
\begin{flushleft}
\quad\quad\quad\large{F}
\end{flushleft}
\end{minipage}

\vspace{-8mm}
\subfloat{\makebox[0.085\textwidth]{}}\hspace{-0.025\textwidth}
\setcounter{subfigure}{3}
\subfloat{\label{fig:SSITT_rev_subfig}\includegraphics[width=0.3\textwidth]{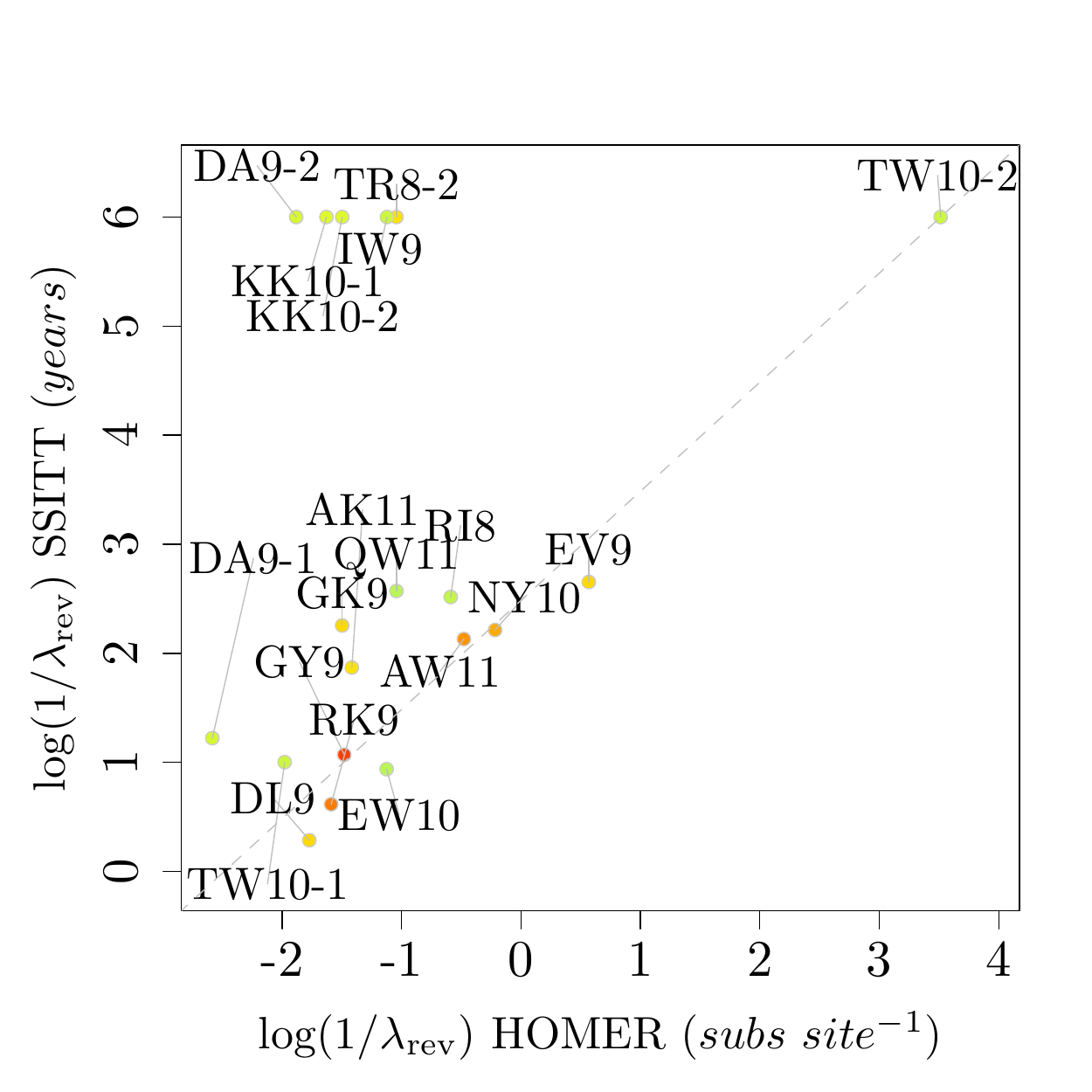}}
\subfloat{\label{fig:3a_geog_rev}\includegraphics[width=0.3\textwidth]{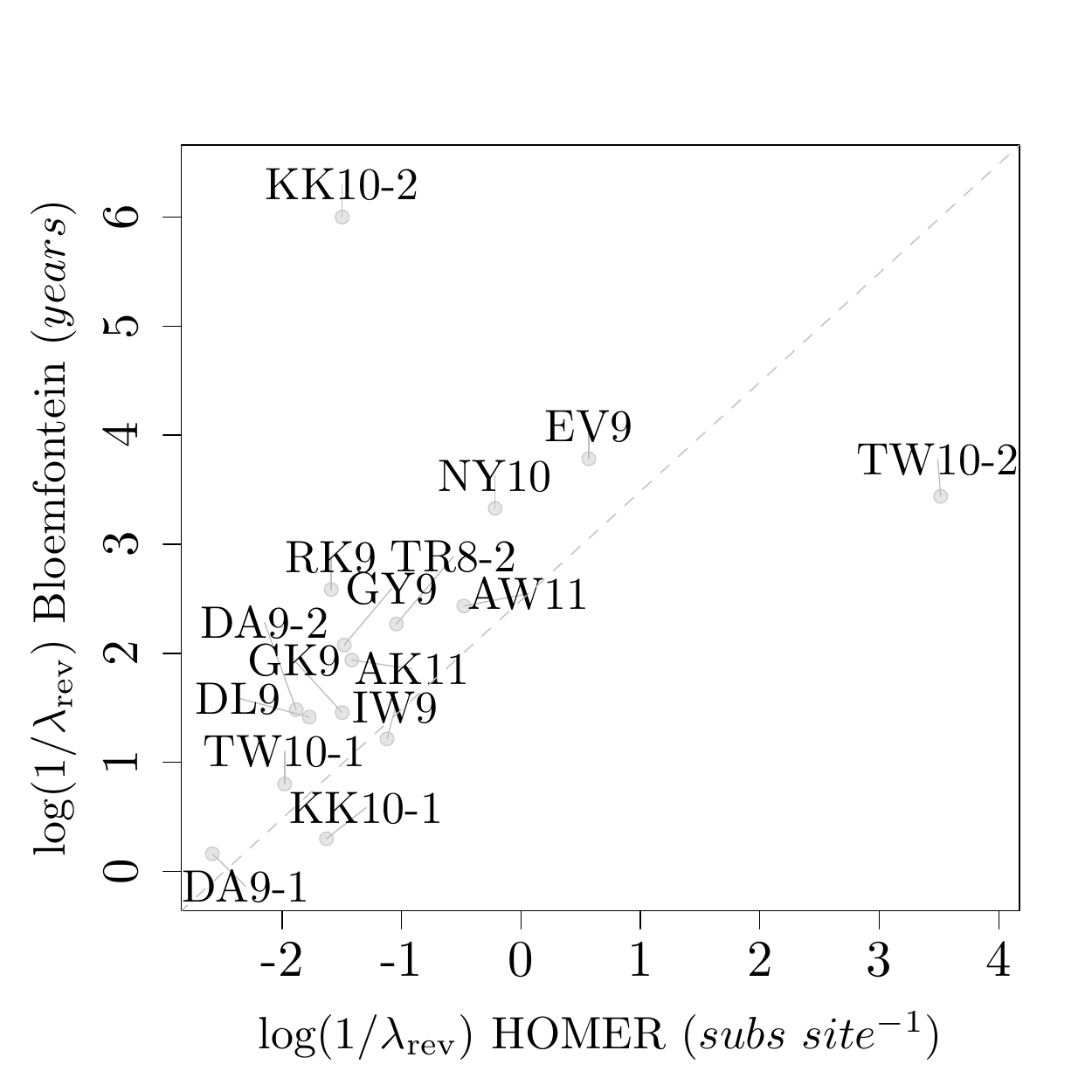}}
\subfloat{\label{fig:4a_geog_rev}\includegraphics[width=0.3\textwidth]{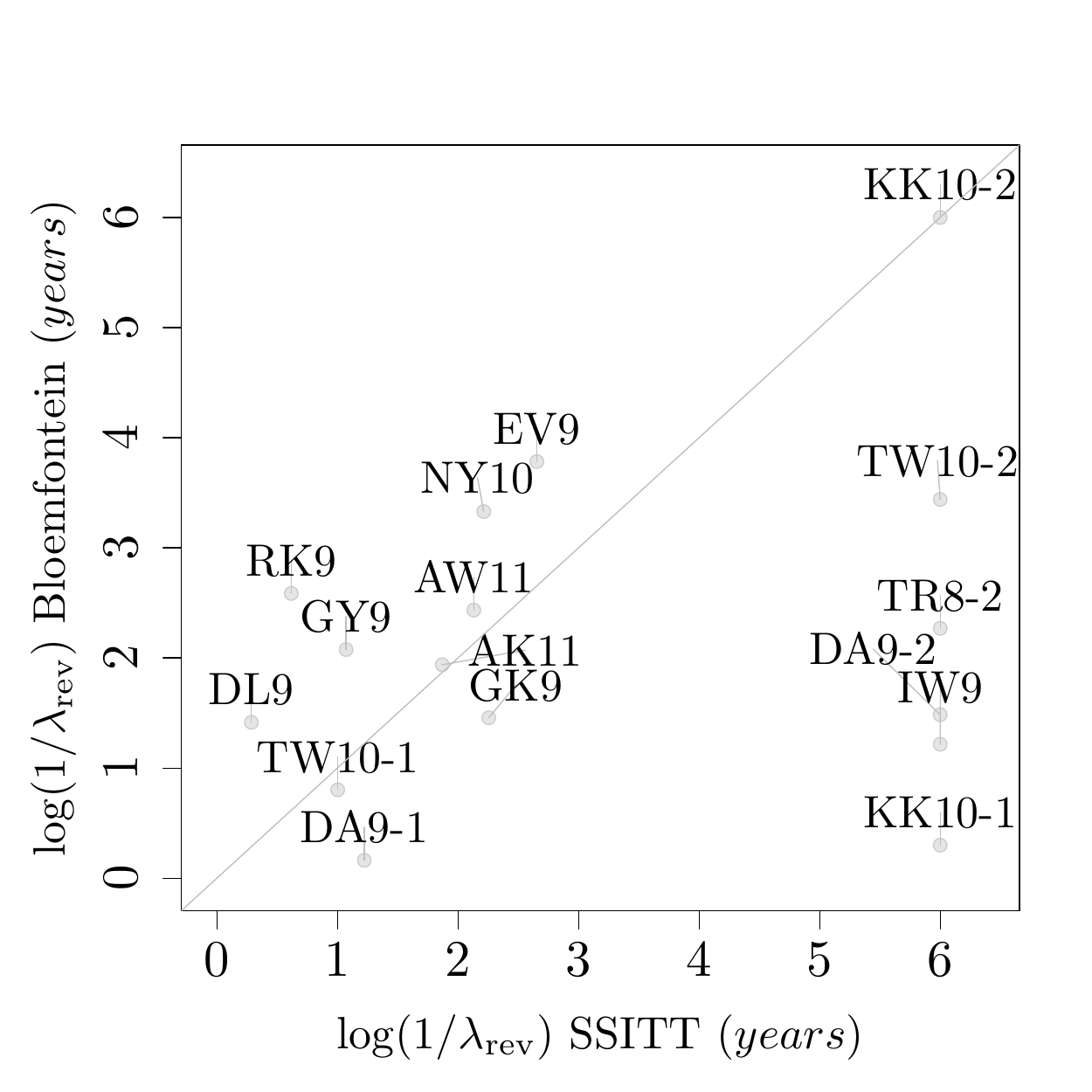}}\\\vspace{-2mm}
\subfloat{\makebox[0.06\textwidth]{}}
\subfloat{\makebox[0.3\textwidth]{$\rho=\geogHOMERSSITTrevrho$; $p=\geogHOMERSSITTrevp$.}}
\subfloat{\makebox[0.3\textwidth]{$\rho=\geogHOMERbloemrevrho$; $p=\geogHOMERbloemrevp$.}}
\subfloat{\makebox[0.3\textwidth]{$\rho=\geogSSITTbloemrevrho$; $p=\geogSSITTbloemrevp$.}}\\
\vspace{3mm}
\caption{{\bf Comparing estimated times to escape/reversion in each pair of cohorts}. Figures \ref{figure:SSITT_HOMER} - \ref{fig:4a_geog} show MAP estimates of time to escape for the SSITT vs. HOMER, Bloemfontein vs. SSITT and Bloemfontein vs. HOMER MAP estimates respectively. Figures \ref{fig:SSITT_rev_subfig} - \ref{fig:4a_geog_rev} show the corresponding MAP estimates of times to reversion. Numbers following a dash refer to the ordering in Tables \ref{SSITT_big_table} - \ref{bloemfontein_big_table}, as some epitopes have more than one escape mutation and/or associated restricting HLA. Negatopes are highlighted in red for time to escape. The scale bar indicates the HLA prevalence taken from the HLA handbook \cite{citeulike:7931760} for each of the restricting HLA types associated to escape in Caucasian populations. $y=x$ is plotted in solid grey where estimates are on the same timescale. Dotted grey lines represent an estimate of $y=x$ after a change of timescale assuming the Swiss and Canadian epidemics are expanding at roughly the same rate.}
\label{figure:HOMER_Bloem_SSITT}
\end{figure*}

\subsubsection*{Escape rates}\label{section:MAPS_across_geogs}
Figure \ref{figure:SSITT_HOMER}-\ref{fig:4a_geog} shows scatter-plots of estimates of average time to escape for the different combinations of populations: HOMER vs. SSITT, HOMER vs. Bloemfontein, and SSITT vs. Bloemfontein respectively. We find strong positive correlations between escape rate estimates for the SSITT and HOMER cohorts ($\rho=\geogHOMERSSITTescrho$; $p=\geogHOMERSSITTescp$). We do not find clustering of restricting HLAs of similar prevalence in any portion of the space of times to escape suggesting that the observed correlation is not an artefact of the HLA prevalence in the two populations. We also observe significant positive correlations between estimates of average time to escape for the remaining population combinations in which HIV-1 subtypes are distinct (analysed SSITT and HOMER viral sequences are subtype B, analysed Bloemfontein viral sequences are subtype C): HOMER vs. Bloemfontein: $\rho=\geogHOMERbloemescrho$; $p=\geogHOMERbloemescp$, and SSITT vs. Bloemfontein: $\rho=\geogSSITTbloemescrho$; $p=\geogSSITTbloemescp$. These results suggest that escape rates are comparable across infected host populations. Under a permutation test (see Methods) we find that the results are significant, $p<0.00001$.

In determining the strength of positive correlation between escape rate estimates across populations, we note that a potential source of bias is the inclusion of negatopes: any mutation at high prevalence across hosts will result in a high escape rate estimate, and consequently may lead to spurious associations. To determine if this is the case in our data we remove negatopes (highlighted in red in Figure \ref{figure:HOMER_Bloem_SSITT}), and re-evaluate correlation coefficients and $p$-values. All positive correlations remained significant after removal of negatopes (HOMER vs. SSITT: $\rho = \geognegremHOMERSSITTescrho$; $p=\geognegremHOMERSSITTescp$, HOMER vs. Bloemfontein: $\rho=\geognegremHOMERbloemescrho$; $p=\geognegremHOMERbloemescp$, and SSITT vs. Bloemfontein: $\rho=\geognegremSSITTbloemescrho$; $p=\geognegremSSITTbloemescp$). The $p$-value for the permutation test to account for multiple testing also remained significant, $p=\pescgeognegmulti$.

We sought to validate our findings with further data. A potential source of error in our Bloemfontein escape rate estimates was the low number of individuals in the Bloemfontein dataset with the restricting HLA (shown in Table \ref{HOMER_big_table}), due to the rarity of these HLA genotypes in African populations coupled with the limited size of the dataset. Additionally, some escape mutations are found at 100\% prevalence or are not present at all in the Bloemfontein dataset. 9/20 rate estimates are obtained from data for which just one individual has the restricting HLA, or 0\%/100\% of HLA matched hosts have the escape mutation. Such escape proportions in HLA matched hosts result in a lack of power to estimate escape rates and can lead to spurious rate estimates \cite{citeulike:12345118} (they may lead to estimates of $\sim$ instantaneous escape or $\sim$ infinite time to escape). To attempt to correct for this source of error we obtained a far larger dataset of African sequences coupled with host HLA genotype data \cite{citeulike:3137, citeulike:4113028, citeulike:12828025, citeulike:9609972, citeulike:12828027, citeulike:12828028, citeulike:12828029, citeulike:12828030}. Unfortunately, such a large dataset was computationally intractable using our integrated model. We therefore turn to the simpler ODE model of Fryer {\it et al.}, noting the agreement between escape rate estimates derived with our integrated method and this compartment based ODE approach \cite{citeulike:12345118}, see Table \ref{table:ODEMAP_table} and Palmer {\it et al} \cite{citeulike:12345118}. The large dataset summarised in Table \ref{table:african_cohort_summary} is not a single study in a single homogeneous mixing population. Nevertheless, we apply the ODE method to all of the data in Table \ref{table:african_cohort_summary}. Scatter-plots of estimates of time to escape using the naive ODE method on the large African dataset against the two HIV-1 subtype B datasets are shown in Figures \ref{figure:HOMER_naive_SSITT}. In both cases we find significant positive correlations ($\rho = \geogHOMERafricaescrho, p = \geogHOMERafricaescp$ and $\rho = \geogSSITTafricaescrho$; $p = \geogSSITTafricaescp$ for African against HOMER and African against SSITT time to escape estimates respectively) which remained significant after removal of negatopes (African vs. HOMER $\rho = \geognegremHOMERafricaescrho$; $p = \geognegremHOMERafricaescp$, African vs. SSITT $\rho = \geognegremSSITTafricaescrho$; $p = \geognegremSSITTafricaescp$). We also note that the strongest disagreement seen in the plots occurs when time to escape is very low in one or other of the populations. This occurs when the escape mutation is at high prevalence in one or other of the populations, possibly as a result of founder effects.
\subsubsection*{Reversion rates}
Figures \ref{fig:SSITT_rev_subfig}-\ref{fig:4a_geog_rev} and \ref{figure:HOMER_SSITT_rev} show the corresponding scatter-plots of estimates of time to reversion. We find a significant agreement between the rank ordering of HOMER and Bloemfontein reversion rate estimates ($\rho = \geogHOMERbloemrevrho, p = \geogHOMERbloemrevp$). This correlation was not repeated in any of the other combination of the analysed cohorts. However, the collection of correlation coefficients observed are significant under a permutation test, $p=\prevgeogmulti$. We note that TLYCVHQ\underline{R}, [\underline{A}]ISPRTLNAW, DRFYK\underline{T}LRA, K\underline{R}WIILGLNK, and KRWII\underline{L}GLNK are estimated to have extremely low reversion rates when using the SSITT data. This is potentially due to the small SSITT cohort size and the inherent difficulty in estimating reversion rates when the restricting HLA type is rare and/or the true underlying reversion rate is low (see simulation studies in \cite{citeulike:12345118}). Further data is required to validate these rate estimates.
\subsection*{Sequence divergence within and between cohorts}
The consistency between escape rate estimates in three different populations seems to suggest that the viral background on which an escape, or pathway to escape is placed is generally a secondary effect during CTL immune evasion. If it weren't, then we would be expect such effects to mask positive correlations between escape rate estimates across host populations. To investigate the extent of viral divergence between the examined datasets, we determined nucleotide and amino acid sequence divergence as measured by Hamming distance (Figure \ref{fig:sequence_divergence}). We evaluate this metric for the region of {\it gag} over which we had the largest number of available sequences. This was to avoid biasing through differences in sequence diversity across {\it gag}. A neighbour joining tree using the K81 model \cite{citeulike:4210163} is shown in Figure \ref{fig:neighbour_joining}. Taken together, Figures \ref{fig:sequence_divergence} and \ref{fig:neighbour_joining} show a dramatic divergence between the subtype B and C sequences, as we would expect. We observe mixing of lineages between Canada and Switzerland, coupled with examples of cohort specific clades.

The consistent and significant positive association between escape rate estimates exists despite a mean divergence of up to $9.2\%$ between viral sequence datasets.
\subsection*{Escape and reversion rates within cohorts vs. {\it in vitro} replicative capacity estimates}\label{section:rates_vs_fitness.}
To examine the importance of single escape variants upon rates of escape and reversion, we consider the most extreme case: complete removal of all sequence variation except at the site of the escape variant. We compare {\it in vitro} replicative capacity of site directed mutants with our escape and reversion estimates. For a collection of 23 epitope variants we had access to sequence data and an estimate of replicative fitness cost from Boutwell {\it et al.}, in which site directed mutagenesis was applied to a subtype B lab strain. If we naively assume an absence of epistasis and that the benefit of evading CTL mediated immunity is equal for all escape variants, then we would expect to see average time to escape negatively correlated with {\it in vitro} replicative capacity, and average time to reversion positively correlated with {\it in vitro} replicative capacity.

\subsubsection*{Subtype B: Escape and reversion rate estimates in SSITT and HOMER datasets vs. {\it in vitro} replicative capacity estimates}

\begin{figure*}
\centering
\begin{minipage}[t]{0.44\textwidth}
\begin{flushleft}
\quad\quad\quad\large{A}
\end{flushleft}
\end{minipage}
\begin{minipage}[t]{0.44\textwidth}
\begin{flushleft}
\quad\quad\quad\large{B}
\end{flushleft}
\end{minipage}
\\\vspace{-10mm}
\subfloat{\label{fig:1a_geog}\includegraphics[page=1,width=0.45\textwidth]{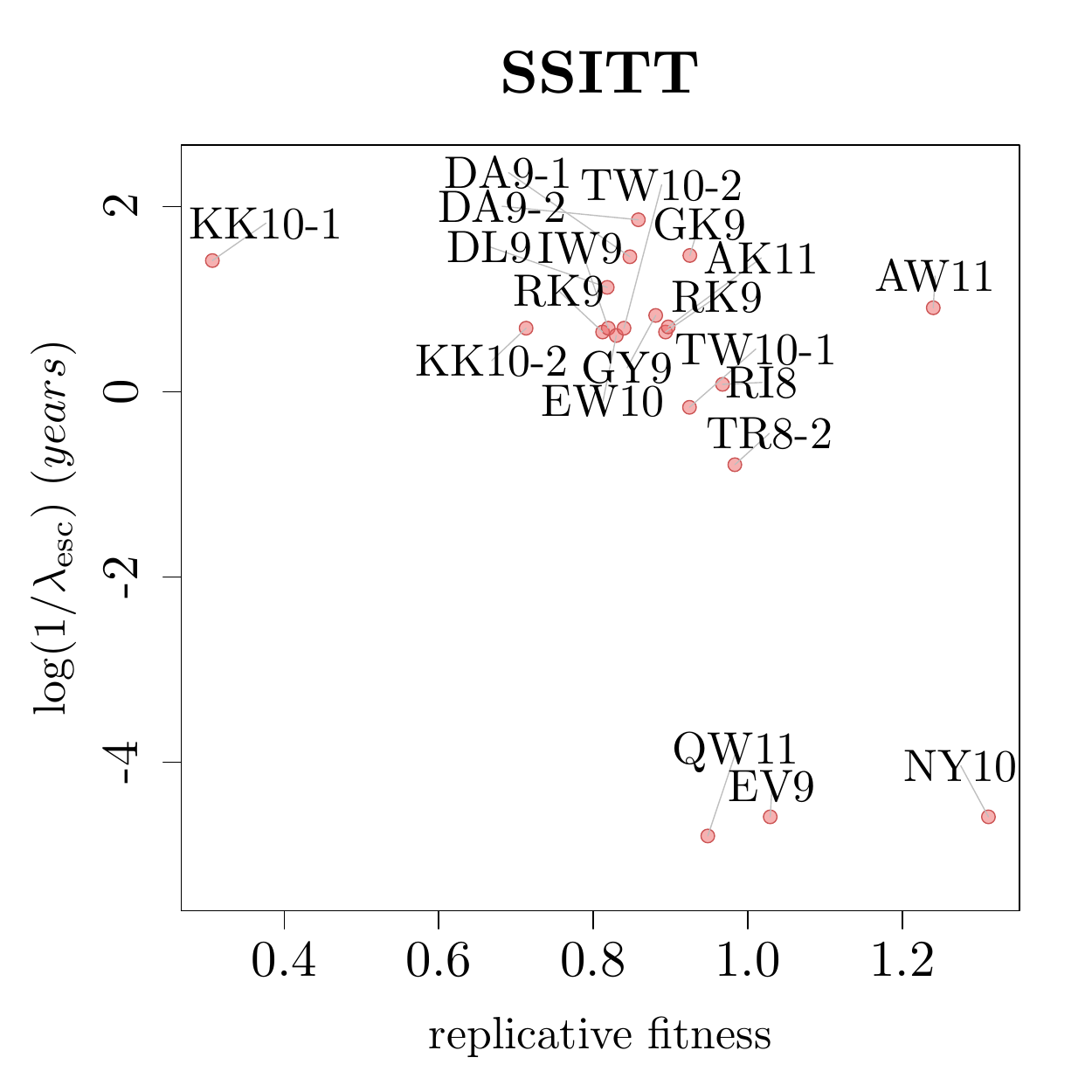}}
\subfloat{\label{fig:1b_geog}\includegraphics[page=1,width=0.45\textwidth]{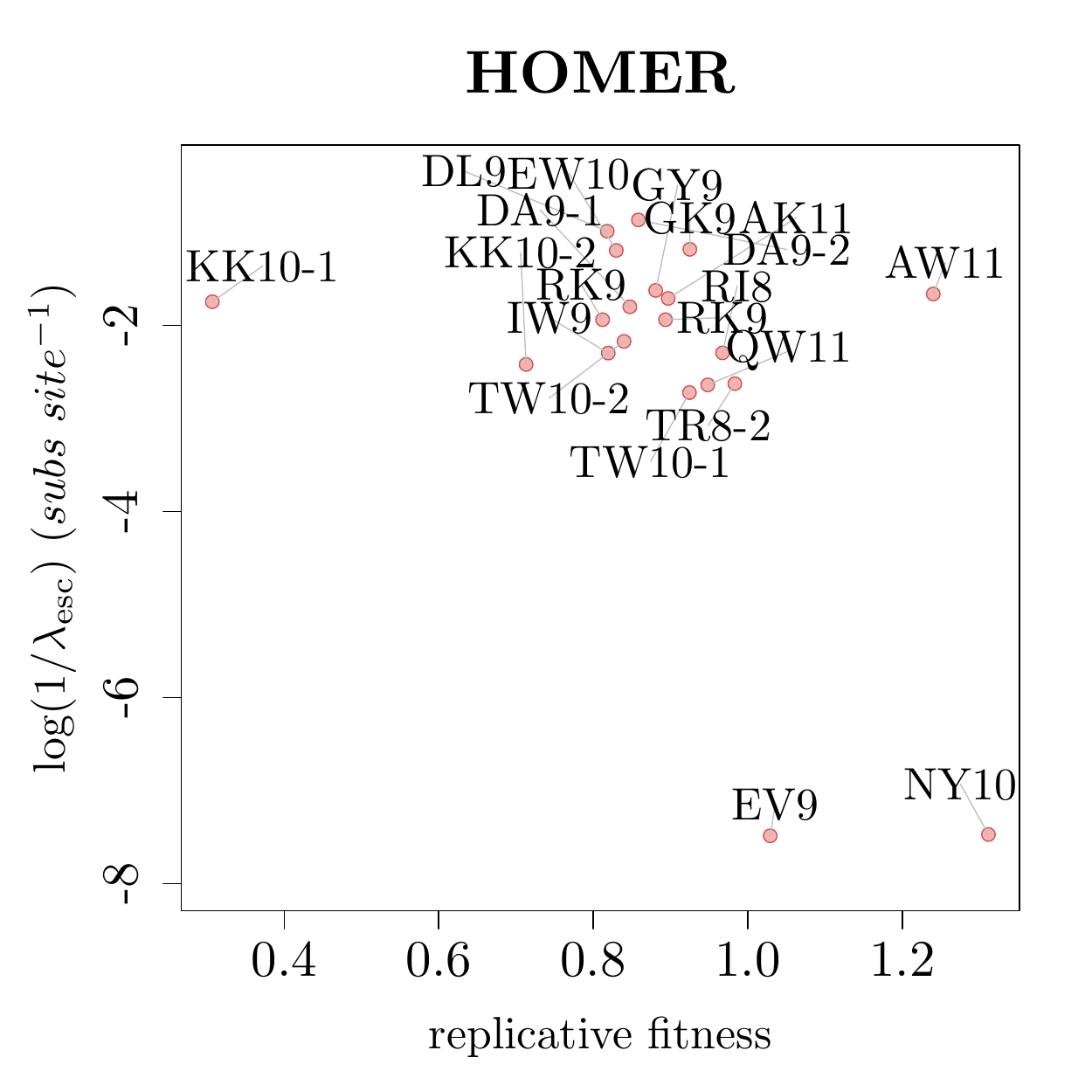}}\\
\begin{minipage}{0.05\textwidth}
\end{minipage}%
\begin{minipage}{0.45\textwidth}
\centering
$\rho=\fitnessSSITTescrho$; $p=\fitnessSSITTescp$.
\end{minipage}%
\begin{minipage}{0.45\textwidth}
\centering
$\rho=\fitnessHOMERescrho$; $p=\fitnessHOMERescp$.
\end{minipage}
\\\vspace{5mm}
\begin{minipage}[t]{0.44\textwidth}
\begin{flushleft}
\quad\quad\quad\large{C}
\end{flushleft}
\end{minipage}
\begin{minipage}[t]{0.44\textwidth}
\begin{flushleft}
\quad\quad\quad\large{D}
\end{flushleft}
\end{minipage}
\\\vspace{-10mm}
\subfloat{\label{fig:2a_geog}\includegraphics[page=2,width=0.45\textwidth]{SSITT_scatter}}
\subfloat{\label{fig:2b_geog}\includegraphics[page=2,width=0.45\textwidth]{HOMER_scatter}}\\
\begin{minipage}{0.05\textwidth}
\end{minipage}%
\begin{minipage}{0.45\textwidth}
\centering
$\rho=\fitnessSSITTrevrho$; $p=\fitnessSSITTrevp$.
\end{minipage}%
\begin{minipage}{0.45\textwidth}
\centering
$\rho=\fitnessHOMERrevrho$; $p=\fitnessHOMERrevp$.
\end{minipage}
\vspace{3mm}
\caption{{\bf Replicative capacity plotted against the MAP estimates of time to escape and reversion}. The first and second columns display estimates obtained using {\it in vivo} data from Switzerland (SSITT) and Canada (HOMER) respectively. Numbers after the epitope abbreviation refer to the ordering in Tables \ref{SSITT_big_table} - \ref{bloemfontein_big_table}, as some epitopes have more than one escape mutation and/or associated restricting HLA. Time to escape is measured in {\it years} except in the case of the HOMER cohort in which it is measured in units of substitutions. Spearman rank correlation coefficients; $\rho$, with associated $p$-values are displayed on each plot.}
\label{figure:scatter_plots}
\end{figure*}
We first examine the correlation between this {\it in vitro} measure of replicative capacity and rate estimates in the two populations in which subtype B is the predominant subtype. Results are displayed in Figure \ref{figure:scatter_plots} (permutation test $p$-value to account for multiple testing of escape rate versus replicative fitness; $p=0.00192$, corresponding $p$-value for reversion; $p=0.0113$). Surprisingly, despite these naive assumptions we indeed observe a negative correlation between time to escape and replicative capacity for the Spearman rank correlation coefficients which reaches significance for the SSITT data ($\rho = \fitnessSSITTescrho$, $p = \fitnessSSITTescp$). We also observe a positive correlation between time to reversion and replicative capacity which reaches significance for the HOMER data ($\rho = \fitnessHOMERrevrho$, $p = \fitnessHOMERrevp$). These findings again suggest that epistatic effects are generally secondary to either the benefits to the virus of CTL escape in HLA matched hosts, or the detriment to viral replicative capacity in HLA mismatched hosts. Such effects appear to be overwhelmed by the impact of insertion or removal of escape mutants. Logically, if epistasis dramatically affected the available pathways to escape/reversion then we would not expect to see time to escape or reversion correlate (negatively and positively respectively) with viral replicative capacity experiments which discard all sequence variation aside from the escape mutant. Finally, we note that outliers in Figure \ref{fig:2a_geog} are exactly those discussed in the previous section. Again we emphasise that larger European sequence datasets with host HLA information should be analysed to determine the accuracy of the reversion rate estimates for these outliers.

\subsubsection*{Subtype C: Escape and reversion rate estimates in the Bloemfontein dataset vs. {\it in vitro} replicative capacity estimates}
We next sought to determine if this correlation was replicated in the Bloemfontein dataset where subtype C predominates. We do not find any such correlation between estimated time to escape/reversion in Bloemfontein (or the full African dataset using the naive ODE method, see Figure \ref{fig:app_fitness_africa}) and {\it in vitro} replicative capacity.  We note that the permutation test $p$-value remains significant after accounting for these further comparisons, $p=\prevfitmulti$. 

\section*{Discussion}

\begin{figure*}
\centering
\includegraphics[width=0.8\textwidth]{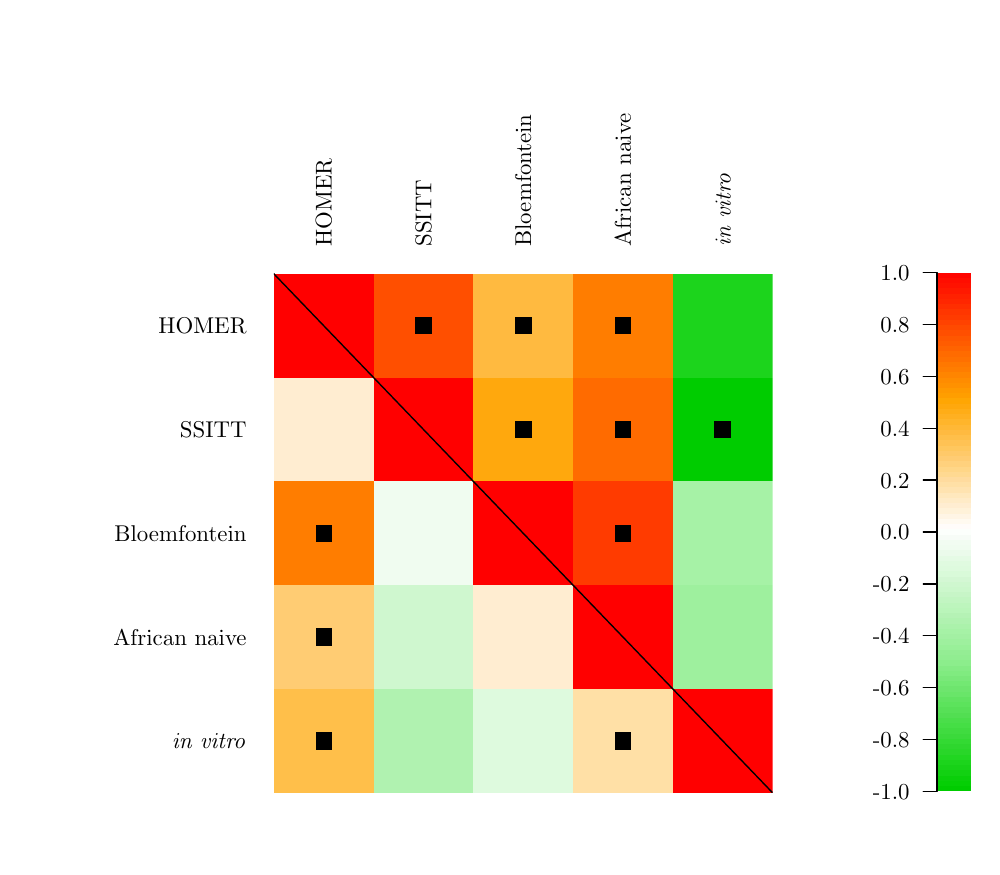}\\
\caption{{\bf Summary of all associations}. Spearman rank correlations between expected time to escape across cohorts and with {\it in vitro} replicative capacity are shown above the leading diagonal. Spearman rank correlations between reversion rates across cohorts and with {\it in vitro} replicative capacity are shown below the leading diagonal. Spearman rank correlation coefficients with $p<0.05$ are highlighted with a black square. Associated Spearman rank correlation coefficients are coloured according to the colourbar.}
\label{figure:summary}
\end{figure*}

There is a large literature detailing the existence and complexity of pathways to escape, compensatory mutations and other forms of epistasis in HIV-1 \cite{citeulike:7807234,citeulike:11396449,citeulike:6544540,citeulike:12198086,citeulike:9219371,citeulike:5427306,citeulike:13799798,citeulike:5342331,citeulike:13801178,citeulike:9443909}. Given this complexity, it is unclear whether the processes of CTL immune escape and reversion can be meaningfully represented by describing each escape variants rate of escape and reversion. In this work, we sought to determine if it makes sense to characterise an escape variants with an escape and reversion rate, and ask if such rate estimates are consistent across host populations. We found significant consistency between expected time to escape for each cohort combination, even across HIV-1 subtypes where average sequence divergence between viral populations was as high as $\sim 9.2\%$. Consistency in escape/reversion rate estimates across populations lead to the following question: is the fitness effect at a single locus a predictor of the tempo of viral escape or reversion? To test this, we considered estimates of {\it in vitro} replicative capacity applied to a subtype B lab strain \cite{citeulike:12198074}. If we naively assume that all escape mutations have the same impact upon CTL immune evasion, then we would expect CTL escape mutations that are more costly {\it in vitro} (i.e. that have a lower viral replicative capacity) to escape more slowly in HLA matched hosts and revert more rapidly in HLA mismatched hosts. The use of {\it in vitro} replicative capacity of single escape variants applied to a lab strain as a measure of its fitness averaged over an entire infected population is clearly a vast simplification. Our method also requires a number of simplifying assumptions \cite{citeulike:12345118}. Despite this, as expected, we find a significant negative correlation between {\it in vitro} replicative capacity and estimated time to escape in the Swiss dataset (SSITT), and a significant positive correlation {\it in vitro} replicative capacity and {\it in vivo} estimates of time until reversion in the HOMER dataset. For the Bloemfontein dataset (consisting of subtype C HIV-1 sequences), no association was found for either escape or reversion estimates. This could suggest a role for epistatic interactions, potentially involving sites distinguishing subtypes B and C. However, we note that the similar replicative capacity of measured escape variants and the noise of such assays, coupled with our low power to estimate many of these rates given the rarity of the associated HLA types in African populations could also lead to a loss of signal of association. Furthermore, founder effects and overlapping epitopes associated to other HLA types have the potential to affect our rate estimates.

Shannon entropy \cite{citeulike:13076300} obtained from viral sequences taken across infected hosts has been used as a surrogate measure of the replicative fitness cost of CTL escape \cite{citeulike:11847211} as it provides a metric for the variability at a given site. Authors thus assume that a site of lower variability will escape more slowly. However, whilst useful, this simple metric ignores the relative prevalence of HLA types in the host population, the dependency structure inherent in the viral genealogy, and has no notion of `escape' in its calculation. By using a simple model of escape and reversion across infected hosts to pick apart the relative contributions to variation observed at escape sites we obtain biologically meaningful parameters which correlate with replicative fitness as we would (perhaps naively) expect. For all of our analysed datasets, Shannon entropy does not correlate with viral replicative capacity ($\rho = 0.193$; $p = 0.208$, $\rho = 0.0226$; $p = 0.462$, $\rho = 0.0850$; $p = 0.361$, for the SSITT, HOMER, and Bloemfontein cohorts respectively).

During the process of escape, it appears that epistatic effects are often overpowered by the effect of a single point mutation; giving rise to consistent escape rate estimates across host populations. This result is powerful as it allows us to inform a null hypothesis when studying a new HIV-1 dataset. Evidence for correlation between reversion rate estimates across populations and with {\it in vitro} replicative capacity is less clear cut. This may be due to the inherent difficulty in estimating reversion rates. Indeed, there is contradiction in the literature regarding the rate of reversion. A few famous epitopes consistently revert in the absence of HLA pressure \cite{citeulike:11847211,citeulike:3176628,citeulike:7807208,citeulike:4113028}. However, whether this readily extends across escape variants is a subject of debate. Some cohort studies have suggested little evidence for reversion \cite{citeulike:13799763,citeulike:11847211,citeulike:13799918}, and cross-sectional estimates suggest a low rate of reversion on average \cite{citeulike:9355716,citeulike:12345118}. However, a large longitudinal study \cite{citeulike:13801164},  transmission pair data \cite{citeulike:13256074} and separate cohort studies \cite{citeulike:7916400}, have suggested that reversion occurs early and is widespread. Other studies have concluded that between 20\% and 30\% of non-synonymous substitutions within host constitute reversions \cite{citeulike:13799788, citeulike:7916400,citeulike:5342321}. Such estimates depend on the authors' definition of reversion, and highlight the need for data sharing (both host HLA and viral sequence data) to critically compare methodologies, increase statistical power to inform estimates and increase our understanding of the underlying biology.

Under our cross-sectional model, reliable reversion rate estimation requires multiple instances of transmission of escape mutations followed by reversion in an HLA mismatched host, an event which occurs less often than transmission of wildtype at the site(s) under consideration (particularly if the restricting HLA is rare, or the escape rate is relatively low). Indeed, simulation studies show that a lack of data and low underlying reversion rates can lead to dramatic underestimation of the true rate \cite{citeulike:12345118}. Despite this, we observe a strong signal of association between the reversion rates estimated using the Bloemfontein data and those estimated using the HOMER dataset. An alternative explanation for the lack of a signal of correlation in reversion rate estimates between cohorts is that epistatic effects are important in a subset of the analysed epitopes. Epistatic effects playing a stronger role in the reversion process does make sense intuitively: on average a viral lineage spends a longer period of time within HLA mismatched hosts (because HLA diversity is so high). Selection to increase replicative capacity acts across the entirety of the viral genome both in the presence and absence of any HLA associated immune pressure (contrast this to CTL associated selective pressure, which is strongest in and around epitopes). The high dimensionality of sequence space would then lead us to expect multiple pathways to a restoration of viral fitness, not all of which may be associated to a simple reversion of the purported escape mutation - particularly if the `survival of the flattest' hypothesis is borne out in reality \cite{citeulike:1573126}.

This study has a number of limitations. Throughout we have applied our methods to subtype B and C viral sequence data from Canada, Switzerland, South Africa, Malawi and Botswana. Access to further host HLA information coupled with cross-sectional HIV-1 sequence data would allow us to determine if similar orderings of escape and reversion rates are observed in further infected host populations, potentially harboring viruses with other HIV-1 subtypes. This will be particularly interesting given the dissimilar escape pathways found by Avila-Rios {\it et al} when analysing a collection of subtype B sequences sampled from Mexico \cite{citeulike:5427306}, and the significant number of further unique HLA associated amino acid variants in other viral subtypes and circulating recombinant forms across the globe \cite{citeulike:13799756}. The escape variants analysed here are all found in {\it gag}. It would therefore be interesting to determine whether our results may be extended to other portions of the HIV-1 genome. Reliably interpreting the effect of replicative capacity on escape and reversion rates in less conserved regions will be difficult. Although replicative fitness assays are improving all the time \cite{citeulike:11611481,citeulike:13798494,citeulike:13798492,citeulike:12198074}, obtaining the required sensitivity to distinguish between small differences in replicative capacity associated to variants in less conserved regions of the viral genome is no easy task. For example, small replicative fitness costs known reverting mutations can often not be distinguished \cite{citeulike:13801178}. The majority of the collection of optimal epitopes studied \cite{Brandereps} have been determined using data from European and North American populations, and thus associated HLA types are likely to be found at high or moderate frequencies within these populations. However, differences in HLA allele frequency distributions across the world often makes rate estimation using data taken from other parts of the world difficult - we sometimes require far larger datasets to obtain rate estimates for escape variants in one region versus another. Datasets with $>\sim 500$ sequences in HIV-1 {\it gag} become intractable using our integrated method. Ability to incorporate vast host/virus datasets would allow us to determine escape and reversion rate estimates for escape variants associated to rare HLA alleles and obtain more accurate estimates for escape mutations associated to more common alleles whilst accounting for dependency structure inherent in viral sequence data.

Our results are summarised in Figure \ref{figure:summary}. Escape rates are consistent across three populations in three different continents and two viral subtypes. We also demonstrate that reversion rates correlate between estimates obtained from a Canadian and African datasets, though further data are required to determine if this observation is true in general.
\section*{Acknowledgements}
We thank Zabrina Brumme and Richard Harrigan for access to host HLA and viral sequence data from the HOMER cohort \cite{citeulike:13587984}.

We thank Christian Boutwell and Todd Allen for sharing raw replicative capacity data used in \cite{citeulike:12198074}.

We thank Bruce Walker for allowing us access to the Durban study cohort \cite{citeulike:9609972, citeulike:12828030}.

\bibliographystyle{pnas_2011}
\bibliography{bibliography,bibliography_manual}
\section*{Supplementary Materials}
\setcounter{figure}{0}
\makeatletter 
\renewcommand{\thefigure}{S\@arabic\c@figure}
\makeatother

\setcounter{table}{0}
\makeatletter 
\renewcommand{\thetable}{S\@arabic\c@table}
\makeatother

\begin{sidewaystable}
\centering
\begin{tabulary}{\textwidth}{l l l l l}
  \toprule
  {\bf Dataset} & Durban \cite{citeulike:9609972, citeulike:12828030} & Mma Bana \cite{citeulike:12828029} & Kimberley \cite{citeulike:12828027} & Thames valley \cite{citeulike:12828027}\\
  {\bf Population analysed} & Durban, South Africa & Gabarone, Botswana & Kimberley, South Africa & \vtop{\hbox{Zimbabwe, South Africa,}\hbox{Malawi and Botswana.}}\\
  {\bf Cohort size} & $n=1218$ & $n=514$ & $n=31$ & $n=102$ \\
  {\bf Sampling date} & 1999 - 2006 & July 2006 - May 2008 & $-$ & $-$ \\
  {\bf Sequences analysed} & $n=929$ & $n=321$ & $n=26$ & \vtop{\hbox{\strut $n=65$}\hbox{Zimbabwe ($n=55$),}\hbox{Malawi ($n=7$)}\hbox{South Africa ($n=3$).}}\\
  {\bf Treatment} & ART naive & ART naive & ART naive & ART naive \\
  {\bf Study requirements} & \vtop{\hbox{\strut Recruited following voluntary}\hbox{counselling and testing in}\hbox{anti-natal or outpatient clinics.}} & Pregnant women & Postnatal mothers & \vtop{\hbox{Southern African subjects}\hbox{attending outpatient HIV}\hbox{clinics in the Thames Valley}\hbox{area of the UK}}\\
\bottomrule
\end{tabulary}
\caption[Summary of four African studies which {\it gag} sequence data and HLA information was available]{Summary of four African studies from which {\it gag} sequence data and HLA information was available. In each case, the number of sequences analysed is lower than the cohort size through a combination of lack of sequencing of {\it gag}, and data cleaning. These data were combined and analysed using the ODE model.}
\label{table:african_cohort_summary}
\end{sidewaystable}
\subsection*{Supplementary methods: Elaborating on step 3 of our inference regime.}\label{subsection:peeling}
After sampling a genealogy from the posterior output by BEAST, we use the host HLA and escape information at the tips of the genealogy to determine estimates of $(\lambda_{\mathrm{esc}}$, $\lambda_{\mathrm{rev}})$. Our algorithm is based on Felsenstein's tree pruning algorithm \cite{citeulike:7890239}, where the tip information is of the form $(a,b)$; $a,b\in\{0,1\}$ such that $\{0,1\}$ in the first entry denotes $\{$HLA mismatch, HLA match$\}$, and $\{0,1\}$ in the second entry denotes $\{$no escape mutation, escape mutation$\}$ at the epitope under investigation. Transmission is assumed to take place at rate $\lambda$. The probability that a given individual is HLA matched at the epitope under investigation is $q$ (assumed to be constant in the population over time, as any selection on hosts would occur on far larger timescales \cite{citeulike:7289645}). We then determine the probability of a given configuration of the data at the tips, conditional on the genealogy, a collection of epidemiological parameters and a collection of parameters $\boldsymbol{\lambda}=\{\lambda_{\mathrm{esc}},\lambda_{\mathrm{rev}},q\}$, which model escape and reversion down each lineage governed by an instantaneous rate matrix $\mathbf{Q}$. The instantaneous rate matrix is based on the following system of ODEs:
\begin{align}
\intertext{HLA mismatched hosts infected with viruses without escape:}
\frac{dY_0^0(t)}{dt}&=\lambda(1-q)\left(Y_0^0(t)+Y_0^1(t)\right)+\lambda_{\mathrm{rev}}Y_1^0(t)-\mu Y_0^0(t)\label{equation_1}\\
\intertext{HLA mismatched hosts infected with viruses with escape:}
\frac{dY_1^0(t)}{dt}&=\lambda(1-q)\left(Y_1^0(t)+Y_1^1(t)\right)-\lambda_{\mathrm{rev}}Y_1^0(t)-\mu Y_1^0(t)\\
\intertext{HLA matched hosts infected with viruses without escape:}
\frac{dY_0^1(t)}{dt}&=\lambda q\left(Y_0^0(t)+Y_0^1(t)\right)-\lambda_{\mathrm{esc}}Y_0^1(t)-\mu Y_0^1(t)\\
\intertext{HLA matched hosts infected with viruses with escape:}
\frac{dY_1^1(t)}{dt}&=\beta cq\left(Y_1^0(t)+Y_1^1(t)\right)+\lambda_{\mathrm{esc}}Y_0^1(t)-\mu Y_1^1(t)\label{equation_4}
\end{align}
where $\lambda$ is the transmission rate. We assume that escape may only occur from the $(1,0)$ state at rate $\lambda_\mathrm{esc}$ and reversion may only occur from the $(0,1)$ state at rate $\lambda_{\mathrm{rev}}$, as in As in \cite{citeulike:9355716,citeulike:12345118}. The instantaneous rate matrix $\mathbf{Q}$ may then be defined as
\begin{align}
\centering
\mathbf{Q} = \bordermatrix{~ & (0,0) & (0,1) & (1,0) & (1,1) \cr
(0,0) & -\hat{\lambda}(t) q & 0 &
  \hat{\lambda}(t)q & 0\cr
(0,1) & \lambda_{\mathrm{rev}} &
  -\lambda_{\mathrm{rev}}-\hat{\lambda}(t) q\ & 0 &
    \hat{\lambda}(t) q \cr
(1,0) & \hat{\lambda}(t)(1-q) & 0 &
    -\lambda_{\mathrm{esc}}-\hat{\lambda}(t)(1-q) & \lambda_{\mathrm{esc}} \cr
(1,1) & 0 & \hat{\lambda}(t)(1-q) & 0 & -\hat{\lambda}(t)(1-q) \cr},
\label{Q_supp}
\end{align}
where $\hat{\lambda}(t)=\lambda p_0(T_{\mathrm{MRCA}}-t)$, with time increasing towards the present. $T_{\mathrm{MRCA}}$ is the time before the present of the most recent common ancestor (MRCA), when $t=0$. $Q_{ij}$ refers to the transition from state $i$ to state $j$. $p_0(t)$ is the probability that a lineage at time $t$ in the past does not have any sampled descendants.
\begin{align}
p_0(t)=1-\frac{\widetilde{\rho}(\lambda-\mu)}{\widetilde{\rho}\lambda+(\lambda(1-\widetilde{\rho})-\mu)\exp(-(\lambda-\mu)t)},
\end{align}
where $\mu$ is the rate of becoming non-infectious, and $\widetilde{\rho}$ is the sampled proportion at the present which we estimate separately \cite{citeulike:2531753}. This extra factor must be included to avoid double counting of transmission events, as we assume transmission events occur at coalescences within the tree. These internal nodes are the transmission events which result in a descendant which is sampled at the present, and occur with probability $1-p_0(T_{MRCA}-t)$. We wish to determine the probability of a branch beginning in state $i$ and ending in state $j$ in order to then evaluate the likelihood over the entire tree. If we define $\mathbf{P}(t)=(P_{(0,0)}(t),P_{(0,1)}(t),P_{(1,0)}(t),P_{(1,1)}(t))$ to be the vector of probabilities of observing each state at time $t$, this amounts to numerically evaluating the solution to
\begin{align}
\dot{\mathbf{P}}(t)=\mathbf{P}(t)\mathbf{Q}(t)
\end{align}
at the end of each branch within the tree, and subject to a prior on the state at the root, and some modifications to ensure a transmission at each internal node, full details of the algorithm are provided in \cite{citeulike:12345118}.

\begin{sidewaystable}
\centering
\begin{tabulary}{\textwidth}{l l l l l l l l l l l}
\toprule
{\small{\bf Epitope}} & {\small{\bf HLA}} & {\small{\bf HLA prop}} &  
            \multicolumn{2}{l}{{\small{\bf with escape}}} & \multicolumn{2}{l}{{\small{\bf Time to escape ($years$)}}} &  
            \multicolumn{2}{l}{{\small{\bf Time to reversion ($years$)}}} &\multicolumn{2}{l}{{\small{\bf Replicative capacity}}} \\ 
            & & & HLA +ve & HLA $-$ve & ODE & MAP & ODE & MAP & rank & actual \\
\midrule
\texttt{GG\underline{K}KKYKLK} & B$^*$08 & $0.143$ & $\sfrac{0}{5}$ & $\sfrac{3}{34}$ & $288$ & $29.6$ & $>10^6$ & $180$ & 8 & $0.925$\\
\texttt{RLRPGGKK\underline{K}} & A$^*$03 & $0.224$ & $\sfrac{7}{11}$ & $\sfrac{4}{28}$ & $3.36$ & $4.41$ & $4.58$ & $4.14$ & 20 & $0.812$\\
\texttt{RLRPGGKK\underline{K}} & A$^*$03 & $0.224$ & $\sfrac{7}{11}$ & $\sfrac{4}{28}$ & $3.36$ & $4.41$ & $4.58$ & $4.14$ & 12 & $0.894$\\
\texttt{GSEELRSL\underline{Y}} & A$^*$01 & $0.262$ & $\sfrac{5}{10}$ & $\sfrac{7}{29}$ & $9.25$ & $6.66$ & $12.7$ & $11.8$ & 13 & $0.881$\\
\texttt{ELRSLYN\underline{T}V*} & B$^*$08 & $0.143$ & $\sfrac{5}{5}$ & $\sfrac{33}{34}$ & $\emph{0.001}$ & $\emph{0.001}$ & $>10^6$ & $451$ & 3 & $1.03$\\
\texttt{[}\texttt{\underline{A}\texttt{]}TLYCVHQR} & A$^*$11 & $0.133$ & $\sfrac{0}{3}$ & $\sfrac{0}{36}$ &$-$ & $-$ & $-$ & $-$ & 4 & $1.02$\\
\texttt{TLYCVHQ\underline{R}} & A$^*$11 & $0.133$ & $\sfrac{3}{3}$ & $\sfrac{23}{36}$ & $0.00257$ & $0.163$ & $50.3$ & $>10^6$ & 5 & $0.983$\\
\texttt{[}\texttt{\underline{A}\texttt{]}ISPRTLNAW} & B$^*$57 & $0.057$ & $\sfrac{1}{7}$ & $\sfrac{8}{31}$ &$26.3$ & $4.85$ & $>10^6$ & $>10^6$ & 9 & $0.819$\\
\texttt{QA\underline{I}SPRTLNAW} & A$^*$25 & $0.042$ & $\sfrac{1}{1}$ & $\sfrac{16}{37}$ & $0.00133$ & $\emph{0.001}$ & $120$ & $373$ & 7 & $0.948$\\
\texttt{K\underline{A}FSPEVIPMF} & B$^*$57 & $0.057$ & $\sfrac{0}{6}$ & $\sfrac{0}{30}$ & $-$ & $-$ & $-$ & $-$ & 22 & $0.387$\\
\texttt{TP\underline{Q}DLNTML} & B$^*$42 & $0.003$ & $\sfrac{0}{0}$ & $\sfrac{0}{57}$ & $-$ & $-$ & $-$ & $-$ & 10 & $0.903$\\
\texttt{ETINEEAA\underline{E}W} & A$^*$25 & $0.042$ & $\sfrac{1}{3}$ & $\sfrac{3}{54}$ & $8.90$ & $4.04$ & $17.4$ & $8.68$ & 17 & $0.830$\\
\texttt{TS\underline{T}LQEQIGW} & B$^*$57 & $0.057$ & $\sfrac{7}{8}$ & $\sfrac{5}{49}$ & $0.610$ & $0.680$ & $8.19$ & $10.1$ & 18 & $0.924$\\
\texttt{TSTLQEQI\underline{G}W} & B$^*$57 & $0.057$ & $\sfrac{4}{8}$ & $\sfrac{12}{49}$ & $7.25$ & $4.85$ & $162$ & $>10^6$ & 16 & $0.840$\\
\texttt{NPPIPVG\underline{E}IY*} & B$^*$35 & $0.196$ & $\sfrac{7}{7}$ & $\sfrac{45}{50}$ & $0.149$ & $\emph{0.001}$ & $215$ & $164$ & 1 & $1.31$\\
\texttt{K\underline{R}WIILGLNK} & B$^*$27 & $0.073$ & $\sfrac{2}{7}$ & $\sfrac{3}{49}$ & $12.2$ & $26.0$ & $13.4$ & $>10^6$ & 23 & $0.307$\\
\texttt{KRWII\underline{L}GLNK} & B$^*$27 & $0.073$ & $\sfrac{5}{7}$ & $\sfrac{7}{49}$ & $1.92$ & $4.85$ & $12.2$ & $>10^6$ & 21 & $0.713$\\
\texttt{RMYSP\underline{T}SI} & B$^*$52 & $0.051$ & $\sfrac{2}{3}$ & $\sfrac{30}{54}$ & $2.59$ & $1.20$ & $>10^6$ & $329$ & 6 & $0.967$\\
\texttt{DRFY\underline{K}TLRA} & B$^*$14 & $0.065$ & $\sfrac{0}{6}$ & $\sfrac{2}{51}$ & $460$ & $28.6$ & $>10^6$ & $16.7$ & 15 & $0.847$\\
\texttt{DRFYK\underline{T}LRA} & B$^*$14 & $0.065$ & $\sfrac{1}{6}$ & $\sfrac{1}{51}$ & $21.5$ & $71.8$ & $7.33$ & $>10^6$ & 14 & $0.858$\\
\texttt{AEQASQ\underline{E}VKNW*} & B$^*$44 & $0.211$ & $\sfrac{11}{13}$ & $\sfrac{26}{44}$ & $2.70$ & $8.06$ & $41.1$ & $135$ & 2 & $1.24$\\
\texttt{DC\underline{K}TILKAL} & B$^*$08 & $0.143$ & $\sfrac{2}{8}$ & $\sfrac{1}{49}$ & $13.6$ & $13.4$ & $2.21$ & $1.93$ & 19 & $0.818$\\
\texttt{ACQGVGGP\underline{G}HK} & A$^*$11 & $0.133$ & $\sfrac{4}{7}$ & $\sfrac{13}{50}$ & $5.62$ & $5.01$ & $23.2$ & $74.1$ & 11 & $0.897$\\
\bottomrule
\end{tabulary}
\caption{{\bf Fitness costs of epitopes in Gag: SSITT}. MAP and ODE \emph{in vivo} estimates of escape and reversion rates for the collection of epitopes in \cite{citeulike:12198074}, obtained using sequence data from the SSITT cohort \cite{citeulike:9530950, citeulike:9405209}. Proportions of individuals possessing each HLA type are taken from the HLA factsbook \cite{citeulike:7931760}. HLA +ve hosts are those with the restricting HLA type. Time to escape and reversion are estimated using the model developed in \cite{citeulike:12345118}. Estimates in italics lie on the boundary of our uniform prior. The lack of an estimate occurs when none of the sampled individuals possessed the restricting HLA allele. Putative subtype B negatopes are labelled with asterisks. Amino acids with square brackets lie outside of the epitope.  Replicative capacity rank is the ranking of the replicative capacity of the escape mutant (where 1 is the mutant strain with the highest replicative capacity). Replicative capacity actual is the measured scaling of replicative capacity relative to a subtype B lab strain \cite{citeulike:12198074}.}\label{SSITT_big_table}
\end{sidewaystable}
\begin{sidewaystable}
\centering
\begin{tabulary}{\textwidth}{l l l l l l l l l l l}
\toprule
{\small{\bf Epitope}} & {\small{\bf HLA}} & {\small{\bf HLA prop}} &  
            \multicolumn{2}{l}{{\small{\bf with escape}}} & \multicolumn{2}{l}{{\small{\bf Time to escape ($subs$ $site^{-1}$)}}} &  
            \multicolumn{2}{l}{{\small{\bf Time to reversion ($subs$ $site^{-1}$)}}} &\multicolumn{2}{l}{{\small{\bf Replicative capacity}}} \\ 
            & & & HLA +ve & HLA $-$ve & ODE & MAP & ODE & MAP & rank & actual \\
\midrule
\texttt{GG\underline{K}KKYKLK} & B$^*$08 & $0.143$ & $\sfrac{2}{21}$ & $\sfrac{9}{140}$ & $0.327$ & $0.0657$ & $>10^4$ & $0.0318$ & 8 & $0.925$\\
\texttt{RLRPGGKK\underline{K}} & A$^*$03 & $0.224$ & $\sfrac{26}{35}$ & $\sfrac{28}{124}$ & $0.00743$ & $0.0115$ & $0.0226$ & $0.0257$ & 20 & $0.812$\\
\texttt{RLRPGGKK\underline{K}} & A$^*$03 & $0.224$ & $\sfrac{26}{35}$ & $\sfrac{28}{124}$ & $0.00743$ & $0.0115$ & $0.0226$ & $0.0257$ & 12 & $0.894$\\
\texttt{GSEELRSL\underline{Y}} & A$^*$01 & $0.262$ & $\sfrac{18}{38}$ & $\sfrac{30}{122}$ & $0.0359$ & $0.0238$ & $0.0495$ & $0.0331$ & 13 & $0.881$\\
\texttt{ELRSLYN\underline{T}V*} & B$^*$08 & $0.143$ & $\sfrac{23}{23}$ & $\sfrac{149}{152}$ & $\emph{0.00001}$ & $\emph{0.00001}$ & $>10^4$ & $3.70$ & 3 & $1.03$\\
\texttt{[}\texttt{\underline{A}\texttt{]}TLYCVHQR} & A$^*$11 & $0.133$ & $\sfrac{0}{23}$ & $\sfrac{0}{153}$ &$-$ & $-$ & $-$ & $-$ & 4 & $1.02$\\
\texttt{TLYCVHQ\underline{R}} & A$^*$11 & $0.133$ & $\sfrac{16}{19}$ & $\sfrac{72}{131}$ & $0.00707$ & $0.00237$ & $0.256$ & $0.0906$ & 5 & $0.983$\\
\texttt{[}\texttt{\underline{A}\texttt{]}ISPRTLNAW} & B$^*$57 & $0.057$ & $\sfrac{1}{6}$ & $\sfrac{35}{151}$ &$0.0722$ & $0.00504$ & $>10^4$ & $0.0755$ & 9 & $0.819$\\
\texttt{QA\underline{I}SPRTLNAW} & A$^*$25 & $0.042$ & $\sfrac{3}{5}$ & $\sfrac{28}{161}$ & $0.0113$ & $0.00229$ & $0.177$ & $0.0906$ & 7 & $0.948$\\
\texttt{K\underline{A}FSPEVIPMF} & B$^*$57 & $0.057$ & $\sfrac{0}{10}$ & $\sfrac{0}{163}$ & $-$ & $-$ & $-$ & $-$ & 22 & $0.387$\\
\texttt{TP\underline{Q}DLNTML} & B$^*$42 & $0.003$ & $\sfrac{0}{0}$ & $\sfrac{1}{174}$ & $-$ & $-$ & $-$ & $-$ & 10 & $0.903$\\
\texttt{ETINEEAA\underline{E}W} & A$^*$25 & $0.042$ & $\sfrac{0}{6}$ & $\sfrac{7}{171}$ & $1.37$ & $0.0639$ & $>10^4$ & $0.0749$ & 17 & $0.830$\\
\texttt{TS\underline{T}LQEQIGW} & B$^*$57 & $0.057$ & $\sfrac{7}{8}$ & $\sfrac{7}{165}$ & $0.00185$ & $0.00189$ & $0.0103$ & $0.0105$ & 18 & $0.924$\\
\texttt{TSTLQEQI\underline{G}W} & B$^*$57 & $0.057$ & $\sfrac{5}{8}$ & $\sfrac{29}{157}$ & $0.0104$ & $0.00673$ & $0.102$ & $3270$ & 16 & $0.840$\\
\texttt{NPPIPVG\underline{E}IY*} & B$^*$35 & $0.196$ & $\sfrac{23}{23}$ & $\sfrac{139}{147}$ & $\emph{0.00001}$ & $\emph{0.00001}$ & $>10^4$ & $0.607$ & 1 & $1.31$\\
\texttt{K\underline{R}WIILGLNK} & B$^*$27 & $0.073$ & $\sfrac{5}{14}$ & $\sfrac{9}{164}$ & $0.0266$ & $0.0180$ & $0.0290$ & $0.0235$ & 23 & $0.307$\\
\texttt{KRWII\underline{L}GLNK} & B$^*$27 & $0.073$ & $\sfrac{8}{10}$ & $\sfrac{18}{148}$ & $0.00368$ & $0.00381$ & $0.0286$ & $0.0318$ & 21 & $0.713$\\
\texttt{RMYSP\underline{T}SI} & B$^*$52 & $0.051$ & $\sfrac{1}{2}$ & $\sfrac{30}{163}$ & $0.0190$ & $0.00506$ & $0.293$ & $0.258$ & 6 & $0.967$\\
\texttt{DRFY\underline{K}TLRA} & B$^*$14 & $0.065$ & $\sfrac{5}{12}$ & $\sfrac{1}{158}$ & $0.0176$ & $0.0158$ & $0.00276$ & $0.00260$ & 15 & $0.847$\\
\texttt{DRFYK\underline{T}LRA} & B$^*$14 & $0.065$ & $\sfrac{1}{14}$ & $\sfrac{2}{159}$ & $0.195$ & $0.136$ & $0.0392$ & $0.0131$ & 14 & $0.858$\\
\texttt{AEQASQ\underline{E}VKNW*} & B$^*$44 & $0.211$ & $\sfrac{25}{26}$ & $\sfrac{97}{126}$ & $0.00292$ & $0.0218$ & $0.276$ & $0.332$ & 2 & $1.24$\\
\texttt{DC\underline{K}TILKAL} & B$^*$08 & $0.143$ & $\sfrac{1}{21}$ & $\sfrac{5}{139}$ & $0.653$ & $0.103$ & $>10^4$ & $0.0169$ & 19 & $0.818$\\
\texttt{ACQGVGGP\underline{G}HK} & A$^*$11 & $0.133$ & $\sfrac{7}{21}$ & $\sfrac{26}{142}$ & $0.0565$ & $0.0195$ & $0.167$ & $0.0383$ & 11 & $0.897$\\
\bottomrule
\end{tabulary}
\caption{{\bf Fitness costs of epitopes in Gag: HOMER}. MAP and ODE \emph{in vivo} estimates of escape and reversion rates for the collection of epitopes in \cite{citeulike:12198074}, obtained using sequence data from the HOMER cohort \cite{citeulike:13587978}. Proportions of individuals possessing each HLA type are taken from the HLA factsbook \cite{citeulike:7931760}. HLA +ve hosts are those with the restricting HLA type. Time to escape and reversion are estimated using the model developed in \cite{citeulike:12345118}. Estimates in italics lie on the boundary of our uniform prior. The lack of an estimate occurs when none of the sampled individuals possessed the restricting HLA allele. Putative subtype B negatopes are labelled with asterisks. Amino acids with square brackets lie outside of the epitope.  Replicative capacity rank is the ranking of the replicative capacity of the escape mutant (where 1 is the mutant strain with the highest replicative capacity). Replicative capacity actual is the measured scaling of replicative capacity relative to a subtype B lab strain in  \cite{citeulike:12198074}.}\label{HOMER_big_table}
\end{sidewaystable}
\begin{sidewaystable}
\centering
\begin{tabulary}{\textwidth}{l l l l l l l l l l l}
\toprule
{\small{\bf Epitope}} & {\small{\bf HLA}} & {\small{\bf HLA prop}} &  
            \multicolumn{2}{l}{{\small{\bf with escape}}} & \multicolumn{2}{l}{{\small{\bf Time to escape ($years$)}}} &  
            \multicolumn{2}{l}{{\small{\bf Time to reversion ($years$)}}} &\multicolumn{2}{l}{{\small{\bf Replicative capacity}}} \\ 
            & & & HLA +ve & HLA $-$ve & ODE & MAP & ODE & MAP & rank & actual \\
\midrule
\texttt{GG\underline{K}KKYKLK} & B$^*$08 & $0.094$ & $\sfrac{1}{13}$ & $\sfrac{8}{190}$ & $101$ & $57.5$ & $60.1$ & $28.6$ & 8 & $0.925$\\
\texttt{RLRPGGKK\underline{K}} & A$^*$03 & $0.125$ & $\sfrac{15}{15}$ & $\sfrac{150}{165}$ & $0.892$ & $\emph{0.001}$ & $>10^6$ & $385$ & 20 & $0.812$\\
\texttt{RLRPGGKK\underline{K}} & A$^*$03 & $0.125$ & $\sfrac{15}{15}$ & $\sfrac{150}{165}$ & $0.892$ & $\emph{0.001}$ & $>10^6$ & $385$ & 12 & $0.894$\\
\texttt{GSEELRSL\underline{Y}} & A$^*$01 & $0.095$ & $\sfrac{12}{18}$ & $\sfrac{60}{180}$ & $3.99$ & $7.10$ & $40.8$ & $119$ & 13 & $0.881$\\
\texttt{ELRSLYN\underline{T}V*} & B$^*$08 & $0.094$ & $\sfrac{13}{13}$ & $\sfrac{190}{191}$ & $\emph{0.001}$ & $\emph{0.001}$ & $>10^6$ & $6070$ & 3 & $1.03$\\
\texttt{[}\texttt{\underline{A}\texttt{]}TLYCVHQR} & A$^*$11 & $0.029$ & $\sfrac{0}{1}$ & $\sfrac{5}{201}$ &$698$ & $67.4$ & $>10^6$ & $>10^6$ & 4 & $1.02$\\
\texttt{TLYCVHQ\underline{R}} & A$^*$11 & $0.029$ & $\sfrac{1}{1}$ & $\sfrac{156}{196}$ & $\emph{0.001}$ & $\emph{0.001}$ & $>10^6$ & $186$ & 5 & $0.983$\\
\texttt{[}\texttt{\underline{A}\texttt{]}ISPRTLNAW} & B$^*$57 & $0.078$ & $\sfrac{8}{9}$ & $\sfrac{39}{190}$ &$0.639$ & $0.843$ & $14.0$ & $16.4$ & 9 & $0.819$\\
\texttt{QA\underline{I}SPRTLNAW} & A$^*$25 & $0.009$ & $\sfrac{0}{0}$ & $\sfrac{51}{198}$ & $-$ & $-$ & $-$ & $-$ & 7 & $0.948$\\
\texttt{K\underline{A}FSPEVIPMF} & B$^*$57 & $0.078$ & $\sfrac{2}{9}$ & $\sfrac{13}{191}$ & $19.8$ & $67.4$ & $20.9$ & $>10^6$ & 22 & $0.387$\\
\texttt{TP\underline{Q}DLNTML} & B$^*$42 & $0.099$ & $\sfrac{13}{41}$ & $\sfrac{17}{151}$ & $13.4$ & $28.6$ & $20.5$ & $84.1$ & 10 & $0.903$\\
\texttt{ETINEEAA\underline{E}W} & A$^*$25 & $0.009$ & $\sfrac{0}{0}$ & $\sfrac{2}{203}$ & $-$ & $-$ & $-$ & $-$ & 17 & $0.830$\\
\texttt{TS\underline{T}LQEQIGW} & B$^*$57 & $0.078$ & $\sfrac{7}{9}$ & $\sfrac{16}{191}$ & $1.26$ & $1.54$ & $5.64$ & $6.35$ & 18 & $0.924$\\
\texttt{TSTLQEQI\underline{G}W} & B$^*$57 & $0.078$ & $\sfrac{9}{9}$ & $\sfrac{187}{189}$ & $\emph{0.001}$ & $\emph{0.001}$ & $>10^6$ & $2750$ & 16 & $0.840$\\
\texttt{NPPIPVG\underline{E}IY*} & B$^*$35 & $0.108$ & $\sfrac{2}{3}$ & $\sfrac{59}{199}$ & $3.64$ & $12.2$ & $27.4$ & $2130$ & 1 & $1.31$\\
\texttt{K\underline{R}WIILGLNK} & B$^*$27 & $0.029$ & $\sfrac{1}{1}$ & $\sfrac{3}{201}$ & $\emph{0.001}$ & $\emph{0.001}$ & $1.90$ & $2.00$ & 23 & $0.307$\\
\texttt{KRWII\underline{L}GLNK} & B$^*$27 & $0.029$ & $\sfrac{1}{1}$ & $\sfrac{13}{202}$ & $\emph{0.001}$ & $26.9$ & $8.61$ & $>10^6$ & 21 & $0.713$\\
\texttt{RMYSP\underline{T}SI} & B$^*$52 & $0.033$ & $\sfrac{0}{0}$ & $\sfrac{199}{201}$ & $-$ & $-$ & $-$ & $-$ & 6 & $0.967$\\
\texttt{DRFY\underline{K}TLRA} & B$^*$14 & $0.068$ & $\sfrac{4}{19}$ & $\sfrac{1}{183}$ & $15.0$ & $15.7$ & $1.41$ & $1.45$ & 15 & $0.847$\\
\texttt{DRFYK\underline{T}LRA} & B$^*$14 & $0.068$ & $\sfrac{2}{18}$ & $\sfrac{9}{181}$ & $54.5$ & $33.5$ & $51.7$ & $30.5$ & 14 & $0.858$\\
\texttt{AEQASQ\underline{E}VKNW*} & B$^*$44 & $0.112$ & $\sfrac{18}{34}$ & $\sfrac{54}{163}$ & $9.48$ & $12.6$ & $68.2$ & $272$ & 2 & $1.24$\\
\texttt{DC\underline{K}TILKAL} & B$^*$08 & $0.094$ & $\sfrac{0}{13}$ & $\sfrac{3}{186}$ & $3060$ & $164$ & $>10^6$ & $26.0$ & 19 & $0.818$\\
\texttt{ACQGVGGP\underline{G}HK} & A$^*$11 & $0.029$ & $\sfrac{0}{1}$ & $\sfrac{90}{196}$ & $31.3$ & $0.618$ & $>10^6$ & $86.8$ & 11 & $0.897$\\
\bottomrule
\end{tabulary}
\caption{{\bf Fitness costs of epitopes in Gag: Bloemfontein}. MAP and ODE \emph{in vivo} estimates of escape and reversion rates for the collection of epitopes in \cite{citeulike:12198074}, obtained using sequence data from the Bloemfontein cohort \cite{citeulike:12828028,citeulike:9628954}. Proportions of individuals possessing each HLA type are taken from the HLA factsbook \cite{citeulike:7931760}. HLA +ve hosts are those with the restricting HLA type. Time to escape and reversion are estimated using the model developed in \cite{citeulike:12345118}. Estimates in italics lie on the boundary of our uniform prior. The lack of an estimate occurs when none of the sampled individuals possessed the restricting HLA allele. Putative subtype B negatopes are labelled with asterisks. Amino acids with square brackets lie outside of the epitope. Replicative capacity rank is the ranking of the replicative capacity of the escape mutant (where 1 is the mutant strain with the highest replicative capacity). Replicative capacity actual is the measured scaling of replicative capacity relative to a subtype B lab strain in  \cite{citeulike:12198074}.}\label{bloemfontein_big_table}
\end{sidewaystable}

\begin{figure*}
\includegraphics[width=0.95\textwidth]{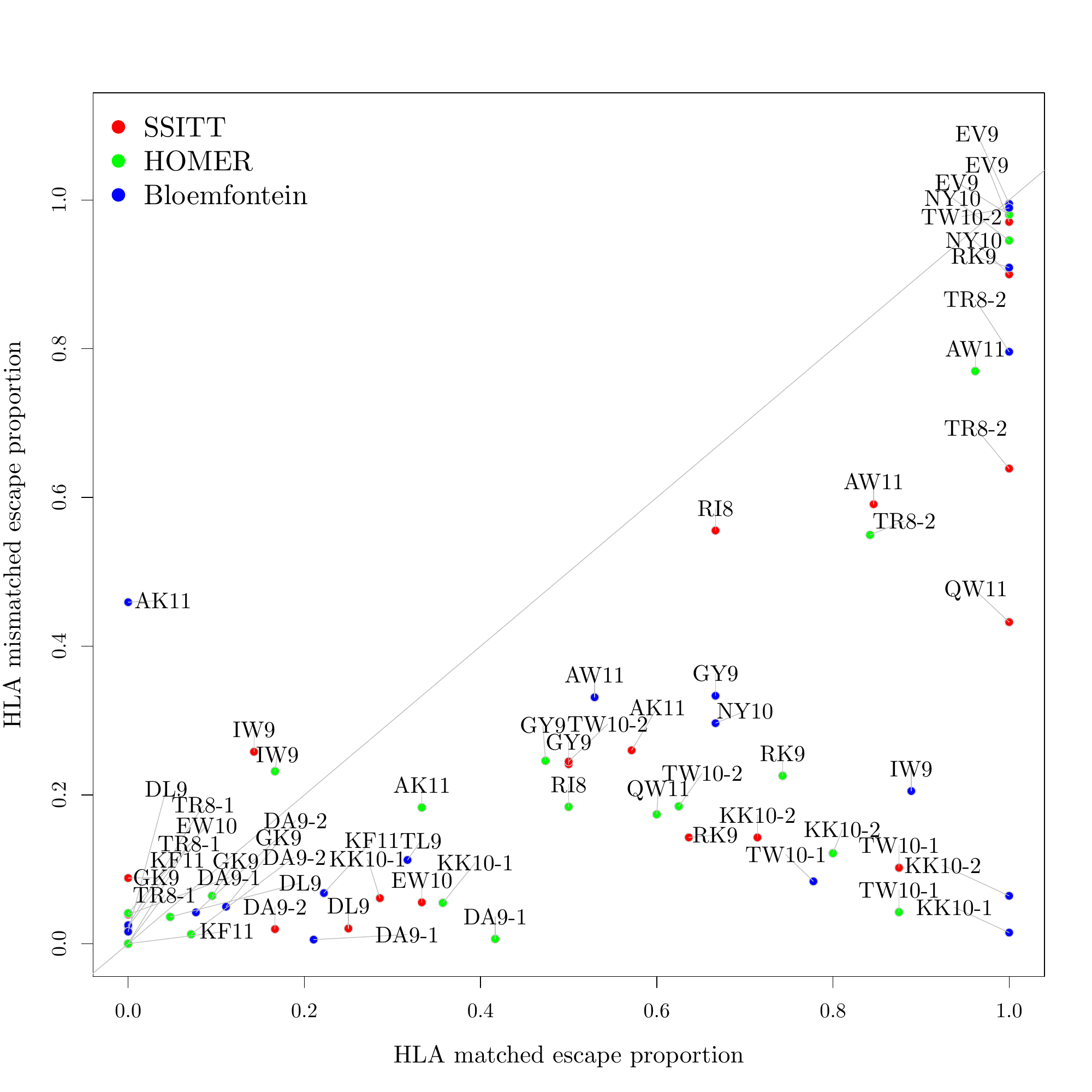}
\caption[Scatterplot of escape proportions in HLA matched and HLA mismatched hosts, coloured by population.]{{\bf Scatter-plot of escape proportions in HLA matched and HLA mismatched hosts, coloured by population}. Data points in which no individuals in a population are HLA matched ($0/0$ escape proportion in HLA matched hosts) were excluded. SSITT, Bloemfontein and HOMER data are shown in red, green and blue respectively. Numbers after the abbreviation of each epitope refers to the ordering in Tables \ref{SSITT_big_table} - \ref{bloemfontein_big_table}, as some epitopes have more than one escape mutation and/or associated restricting HLA.}
\label{figure:HLA_proportions}
\end{figure*}

\begin{table}
\centering
\begin{tabulary}{\textwidth}{l l l l l}
\toprule
      & \multicolumn{2}{c}{Escape} & \multicolumn{2}{c}{Reversion} \\
Study & $\rho$ & $p$ & $\rho$ & $p$\\
\midrule
SSITT & \SSITTODEMAPescrho & \SSITTODEMAPescp & \SSITTODEMAPrevrho & \SSITTODEMAPrevp\\
HOMER & \HOMERODEMAPescrho & \HOMERODEMAPescp & \HOMERODEMAPrevrho & \HOMERODEMAPrevp\\
Bloemfontein & \bloemODEMAPescrho & \bloemODEMAPescp & \bloemODEMAPrevrho & \bloemODEMAPrevp\\
\bottomrule
\end{tabulary}
\caption[Summary table of the Spearman rank correlation coefficient and associated $p$-values between rate estimates using the our integrated method and the ODE method.]{Summary table of the Spearman rank correlation coefficient and associated $p$-values between rate estimates using the our integrated method and the ODE method.}\label{table:ODEMAP_table}
\end{table}

\begin{figure*}
\centering
\begin{minipage}[t]{0.44\textwidth}
\begin{flushleft}
\quad\quad\quad\large{A}
\end{flushleft}
\end{minipage}
\begin{minipage}[t]{0.44\textwidth}
\begin{flushleft}
\quad\quad\quad\large{B}
\end{flushleft}
\end{minipage}
\\\vspace{-10mm}
\subfloat{\label{fig:5a_geog_esc}\includegraphics[width=0.45\textwidth]{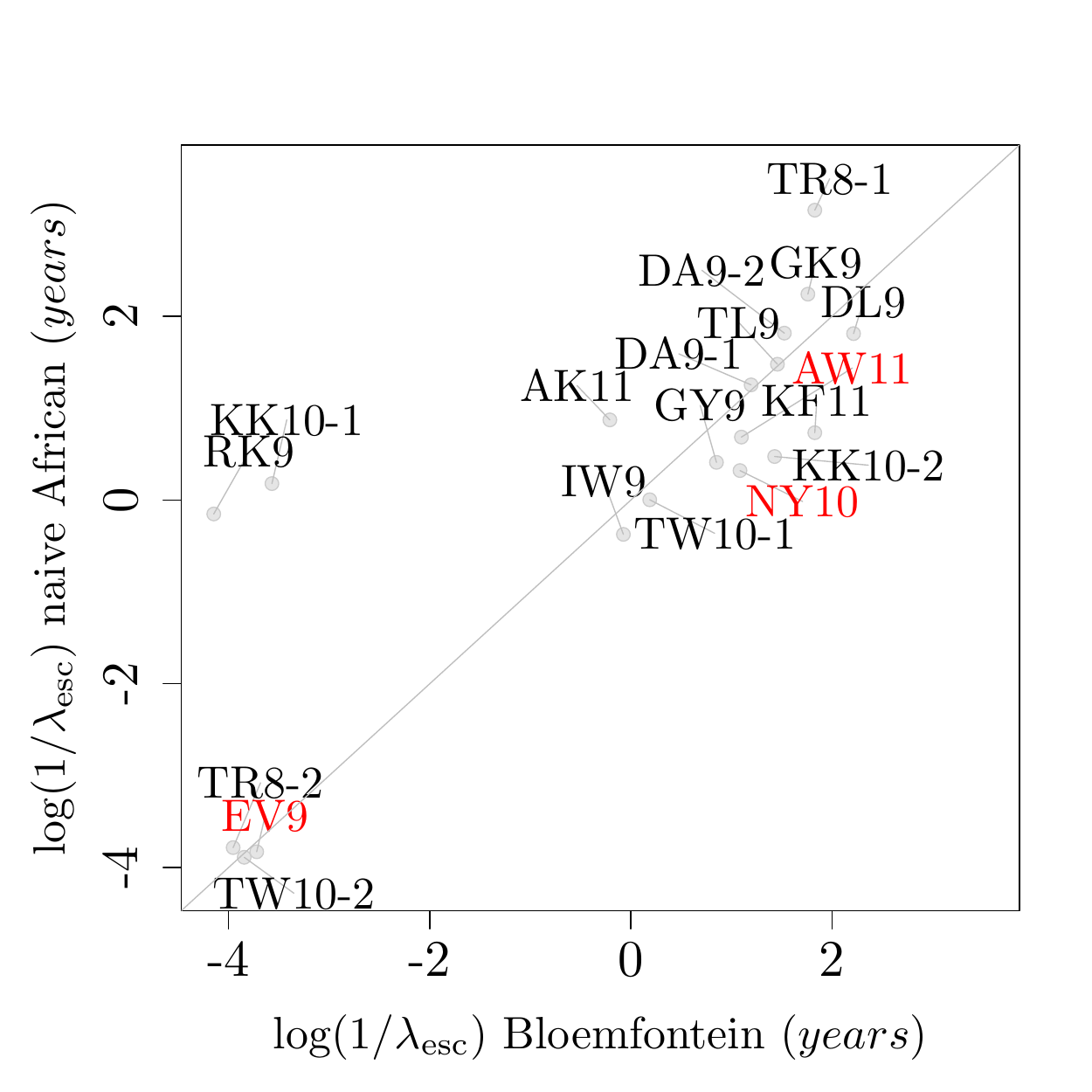}}
\subfloat{\label{fig:5b_geog_rev}\includegraphics[width=0.45\textwidth]{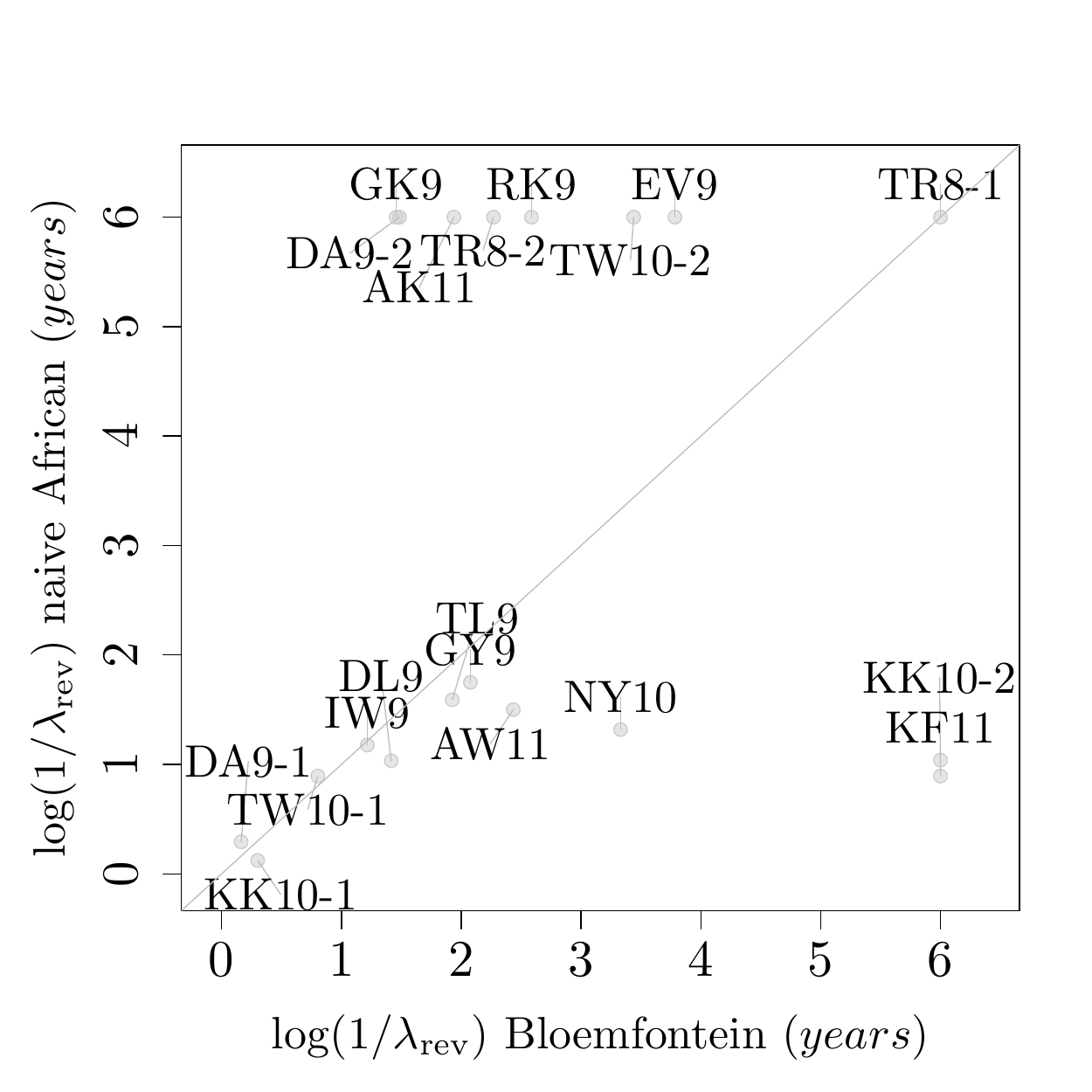}}\\
\subfloat{\makebox[0.45\textwidth]{$\rho=\geogbloemafricaescrho$; $p=\geogbloemafricaescp$.}}
\subfloat{\makebox[0.45\textwidth]{$\rho=\geogbloemafricarevrho$; $p=\geogbloemafricarevp$.}}\\\vspace{-2mm}
\subfloat{\makebox[0.45\textwidth]{}}
\vspace{3mm}
\caption{{\bf Comparing time to escape/reversion as estimated from the Bloemfontein data using the integrated method to estimates generated using all available African data and the naive approach}. Spearman rank correlation coefficients ($\rho$) with associated $p$-values are displayed. Negatopes are highlighted in red for time to escape. Numbers after the abbreviation of each epitope refers to the ordering in Tables \ref{SSITT_big_table} - \ref{bloemfontein_big_table}, as some epitopes have more than one escape mutation and/or associated restricting HLA. $y=x$ is shown in solid grey.}
\label{figure:african_bloem}
\end{figure*}

\begin{figure*}
\centering
\begin{minipage}[t]{0.44\textwidth}
\begin{flushleft}
\quad\quad\quad\large{A}
\end{flushleft}
\end{minipage}
\begin{minipage}[t]{0.44\textwidth}
\begin{flushleft}
\quad\quad\quad\large{B}
\end{flushleft}
\end{minipage}
\\\vspace{-10mm}
\subfloat{\label{fig:3b_geog}\includegraphics[width=0.45\textwidth]{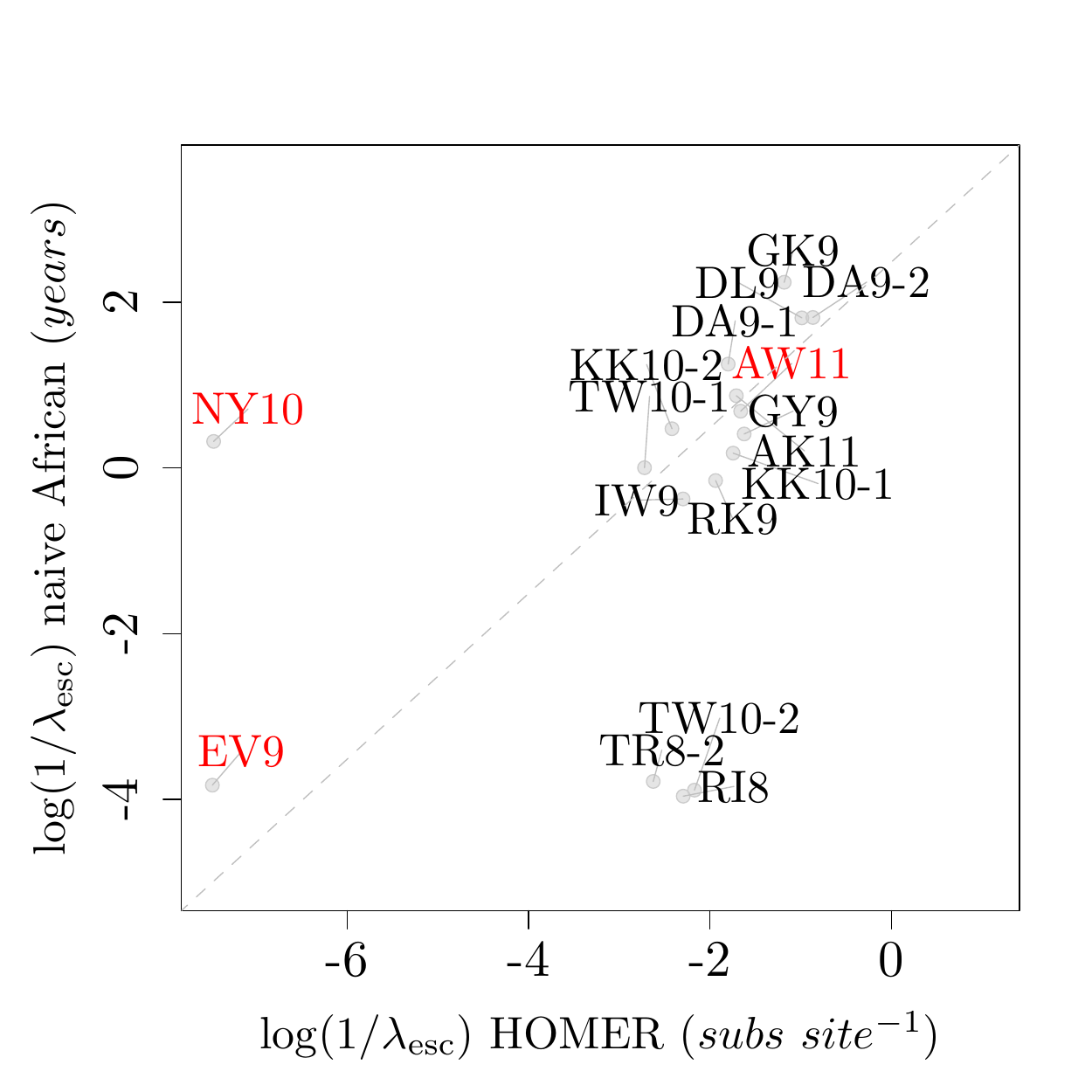}}
\subfloat{\label{fig:4b_geog}\includegraphics[width=0.45\textwidth]{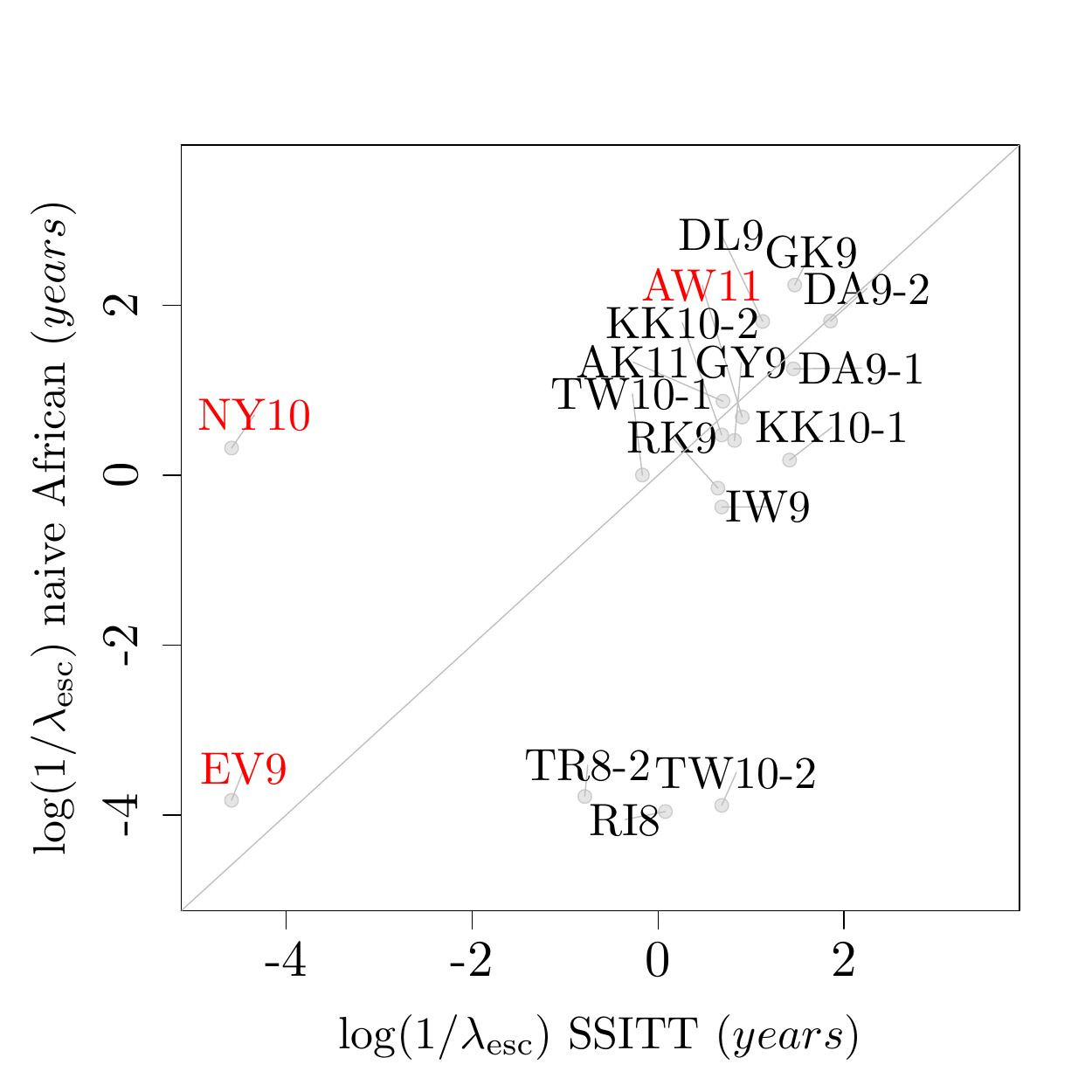}}\\
\subfloat{\makebox[0.45\textwidth]{$\rho=\geogHOMERafricaescrho$; $p=\geogHOMERafricaescp$.}}
\subfloat{\makebox[0.45\textwidth]{$\rho=\geogSSITTafricaescrho$; $p=\geogSSITTafricaescp$.}}\\\vspace{-2mm}
\vspace{3mm}
\caption{{\bf Comparing time to escape/reversion as estimated from the SSITT/HOMER data using the integrated method to estimates generated using all available African data and the naive approach}. In Figures \ref{fig:3b_geog} - \ref{fig:4b_geog} SSITT and HOMER MAP estimates are plotted against a naive estimate of the escape rate. Spearman rank correlation coefficients ($\rho$) with associated $p$-values are displayed. Negatopes are highlighted in red. Numbers after the epitope abbreviation refer to the ordering in Tables \ref{SSITT_big_table} - \ref{bloemfontein_big_table}, as some epitopes have more than one escape mutation and/or associated restricting HLA. We also plot $y=x$ in solid grey where estimates are on the same timescale. Dotted grey lines represent an estimate of $y=x$ after a change of timescale assuming the Swiss and Canadian epidemics are expanding at roughly the same rate.}
\label{figure:HOMER_naive_SSITT}
\end{figure*}

\begin{figure*}
\centering
\begin{minipage}[t]{0.44\textwidth}
\begin{flushleft}
\quad\quad\quad\large{A}
\end{flushleft}
\end{minipage}
\begin{minipage}[t]{0.44\textwidth}
\begin{flushleft}
\quad\quad\quad\large{B}
\end{flushleft}
\end{minipage}
\\\vspace{-10mm}
\subfloat{\label{fig:3b_geog_rev}\includegraphics[width=0.45\textwidth]{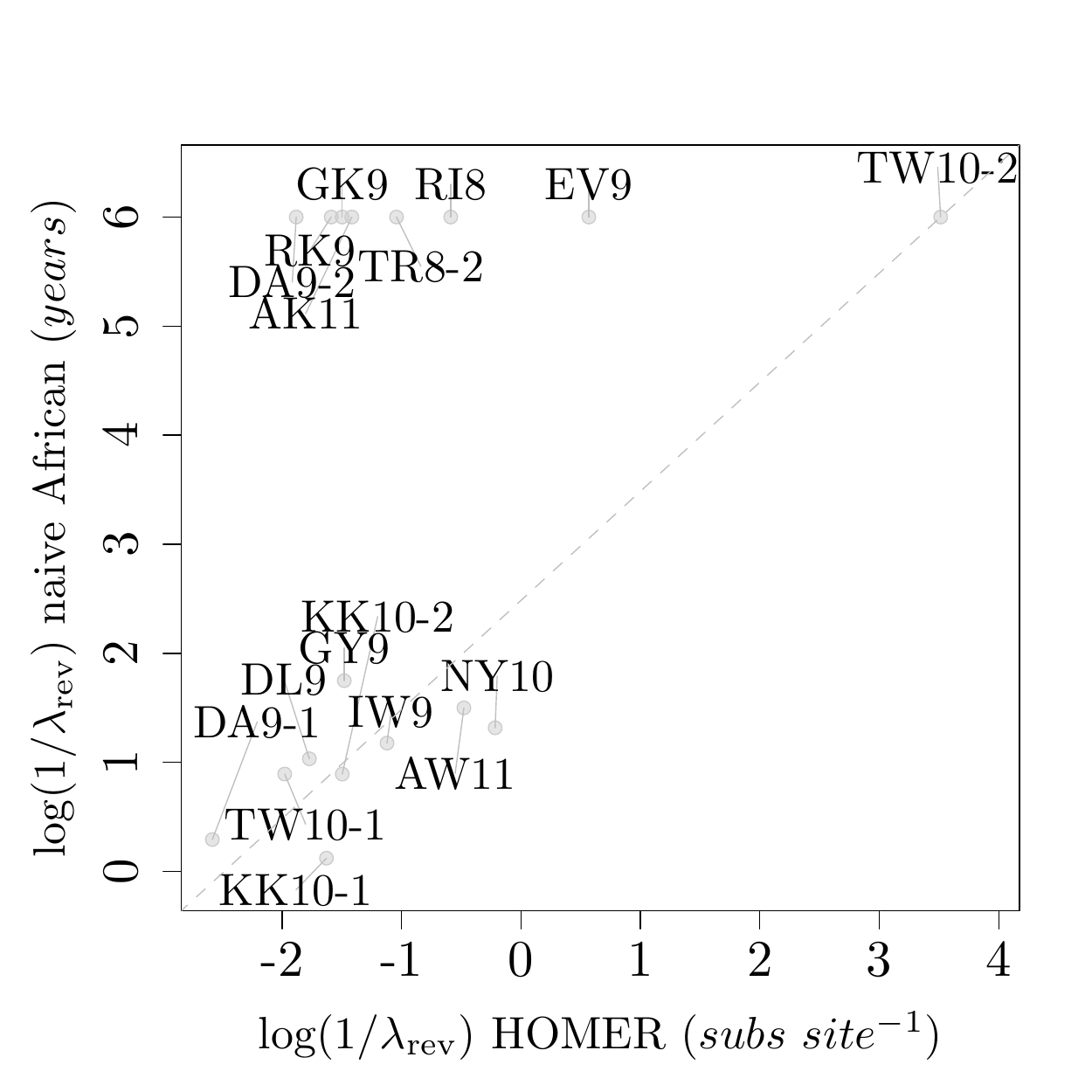}}
\subfloat{\label{fig:4b_geog_rev}\includegraphics[width=0.45\textwidth]{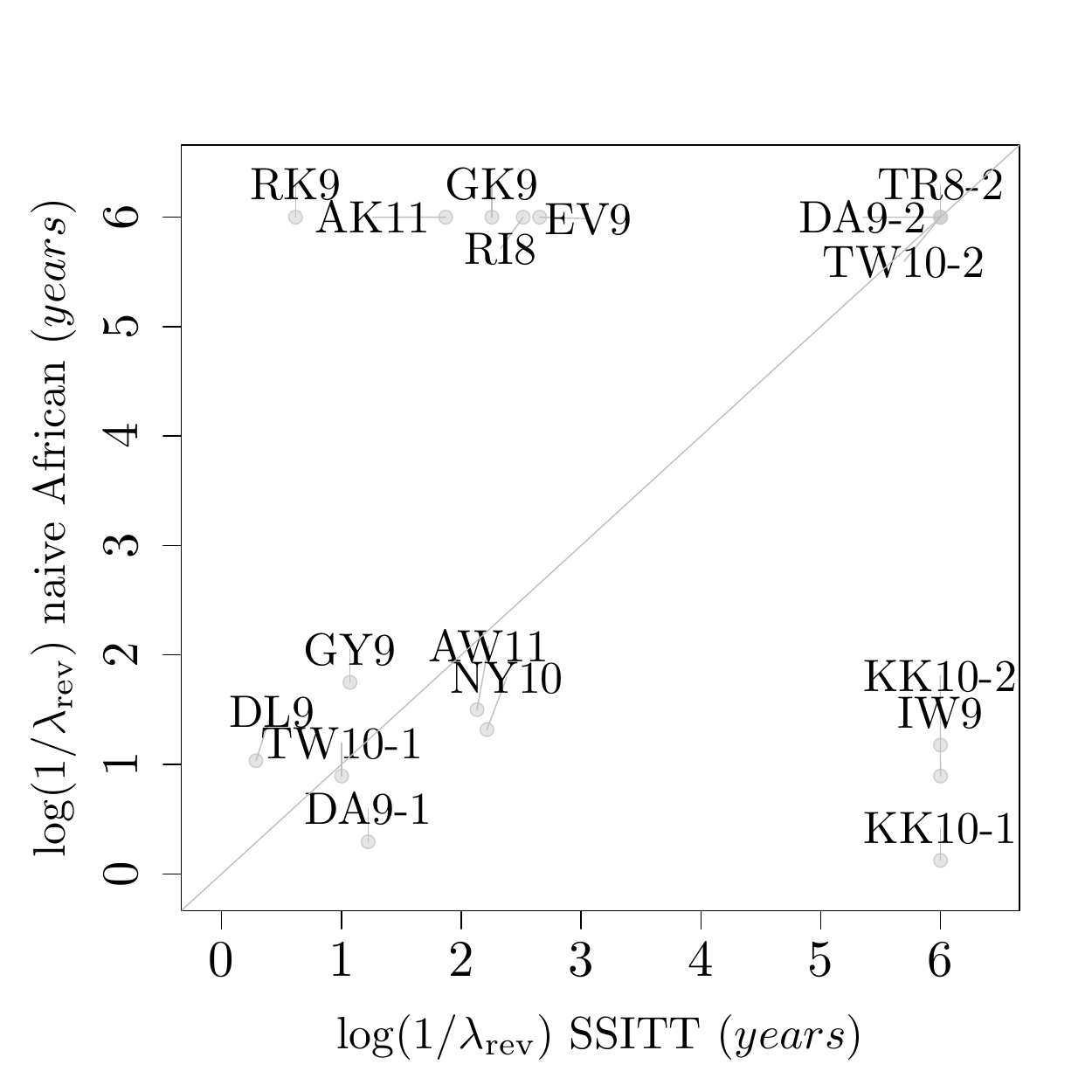}}\\
\subfloat{\makebox[0.45\textwidth]{$\rho=\geogHOMERafricarevrho$; $p=\geogHOMERafricarevp$.}}
\subfloat{\makebox[0.45\textwidth]{$\rho=\geogSSITTafricarevrho$; $p=\geogSSITTafricarevp$.}}
\vspace{3mm}
\caption{{\bf Comparing time to reversion as estimated from the SSITT/HOMER data to estimates of time to reversion using the large African dataset and the naive approach}. In Figures \ref{fig:3b_geog_rev} - \ref{fig:4b_geog_rev} SSITT and HOMER MAP estimates are plotted against a naive African estimate of the reversion rate. Spearman rank correlation coefficients ($\rho$) with associated $p$-values are displayed. Numbers after the abbreviation of each epitope refers to the ordering in Tables \ref{SSITT_big_table} - \ref{bloemfontein_big_table}, as some epitopes have more than one escape mutation and/or associated restricting HLA. $y=x$ is plotted in solid grey where estimates are on the same timescale. Dotted grey lines represent an estimate of $y=x$ after a change of timescale assuming the Swiss and Canadian epidemics are expanding at roughly the same rate.}
\label{figure:HOMER_SSITT_rev}
\end{figure*}

\begin{table}
\centering
\begin{tabulary}{\textwidth}{l l l l l}
\toprule
      & \multicolumn{2}{c}{Escape} & \multicolumn{2}{c}{Reversion} \\
Study & $\rho$ & $p$ & $\rho$ & $p$\\
\midrule
SSITT & \geogSSITTafricaescrho & \geogSSITTafricaescp & \geogSSITTafricarevrho & \geogSSITTafricarevp\\
HOMER & \geogHOMERafricaescrho & \geogHOMERafricaescp & \geogHOMERafricarevrho & \geogHOMERafricarevp\\
Bloemfontein & \geogbloemafricaescrho & \geogbloemafricaescp & \geogbloemafricarevrho & \geogbloemafricarevp\\
\bottomrule
\end{tabulary}
\caption[Summary table of the Spearman rank correlation coefficient and associated $p$-values between escape and reversion rate estimates using the ODE model applied to the large African dataset, and the SSITT, HOMER and Bloemfontein datasets.]{Summary table of the Spearman rank correlation coefficient and associated $p$-values between escape and reversion rate estimates using the ODE model applied to the large African dataset, and the SSITT, HOMER and Bloemfontein datasets.}
\end{table}

\begin{figure*}
\centering
\begin{minipage}[t]{0.44\textwidth}
\begin{flushleft}
\quad\quad\quad\large{A}
\end{flushleft}
\end{minipage}
\begin{minipage}[t]{0.44\textwidth}
\begin{flushleft}
\quad\quad\quad\large{B}
\end{flushleft}
\end{minipage}
\\\vspace{-10mm}
\subfloat{\label{fig:1c_geog}\includegraphics[page=1,width=0.45\textwidth]{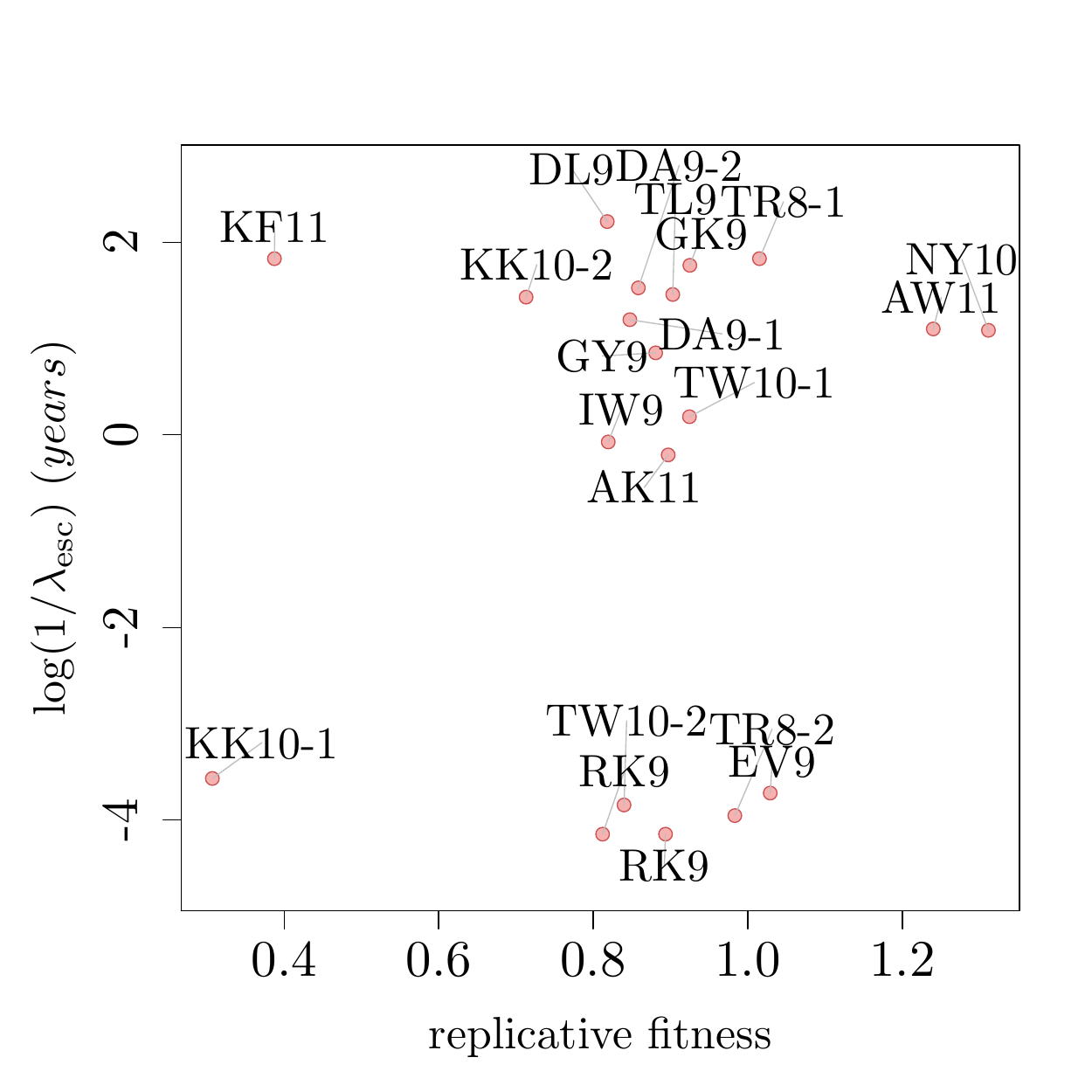}}
\subfloat{\label{fig:2c_geog}\includegraphics[page=2,width=0.45\textwidth]{bloemfontein_scatter}}\\
\subfloat{\makebox[0.45\textwidth]{$\rho=\fitnessbloemescrho$; $p=\fitnessbloemescp$.}}
\subfloat{\makebox[0.45\textwidth]{$\rho=\fitnessbloemrevrho$; $p=\fitnessbloemrevp$.}}
\vspace{3mm}
\caption[Scatter-plots of time to escape and reversion estimates against {\it in vitro} viral
replicative capacity: Bloemfontein.]{{\bf Scatter-plots of time to escape and reversion estimates against {\it in vitro} viral replicative capacity: Bloemfontein}. Replicative capacity plotted against the MAP estimates of time to escape is shown in Figure \ref{fig:1c_geog}, and the MAP estimates of time to reversion in Figure \ref{fig:2c_geog}. Epitopes are labelled by the first three amino acids. Numbers after the abbreviation of each epitope refers to the ordering in Tables \ref{SSITT_big_table} - \ref{bloemfontein_big_table}, as some epitopes have more than one escape mutation and/or associated restricting HLA. Spearman rank correlation coefficients; $\rho$, with associated $p$-values are displayed on each plot.}
\label{figure:scatter_plots_bloem}
\end{figure*}

\begin{figure*}
\centering
\begin{minipage}[t]{0.44\textwidth}
\begin{flushleft}
\quad\quad\quad\large{A}
\end{flushleft}
\end{minipage}
\begin{minipage}[t]{0.44\textwidth}
\begin{flushleft}
\quad\quad\quad\large{B}
\end{flushleft}
\end{minipage}
\\\vspace{-10mm}
\subfloat{\label{figure:naive_african_fitness_esc}\includegraphics[width=0.45\textwidth]{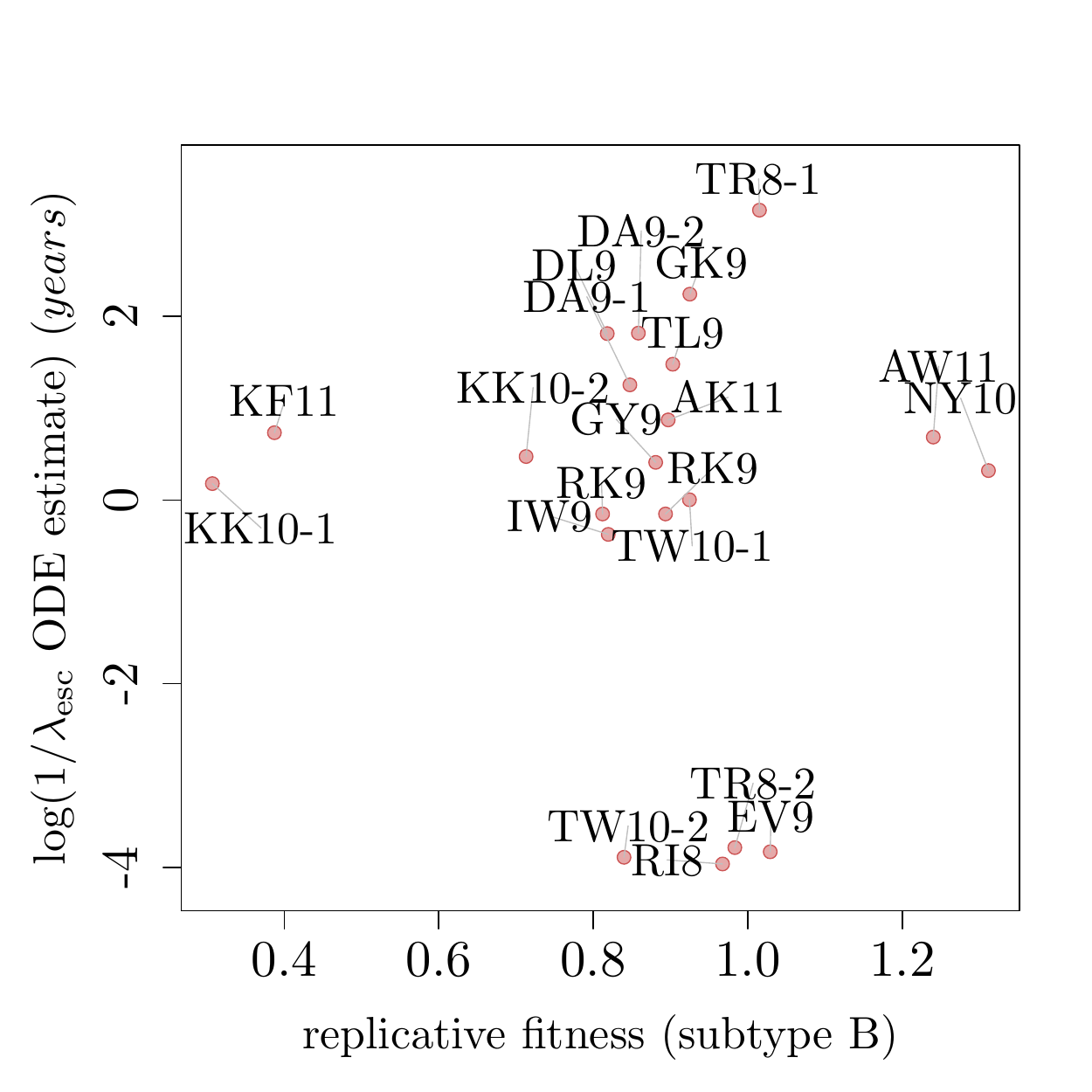}}
\subfloat{\label{figure:naive_african_fitness_rev}\includegraphics[width=0.45\textwidth]{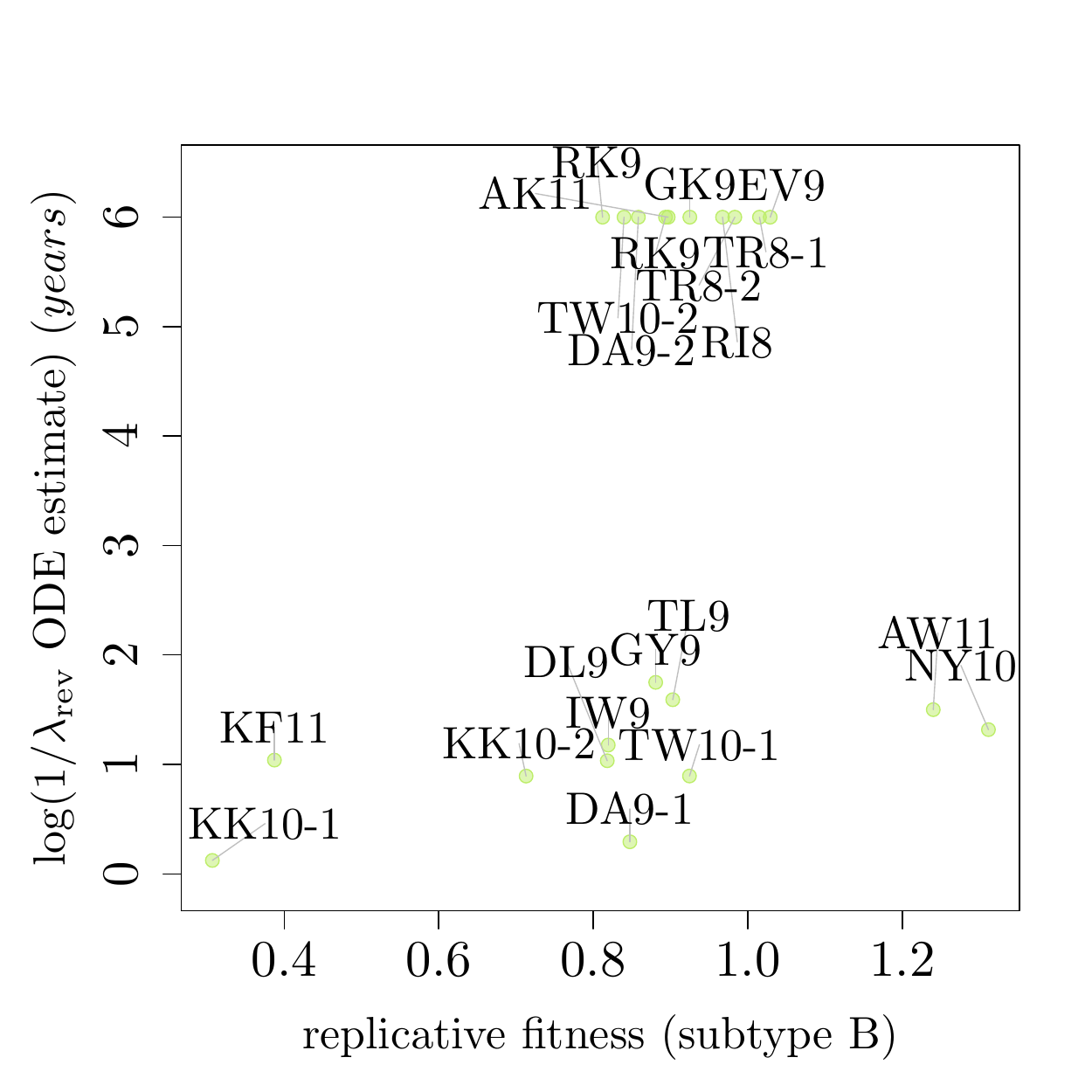}}\\
\subfloat{\makebox[0.45\textwidth]{$\rho=\fitnessafricaescrho$; $p=\fitnessafricaescp$.}}
\subfloat{\makebox[0.45\textwidth]{$\rho=\fitnessafricarevrho$; $p=\fitnessafricarevp$.}}
\vspace{3mm}
\caption{{\bf Scatter-plots of estimates of time to escape and reversion estimated from all African data using naive ODE approach against {\it in vitro} replicative capacity}. Figure \ref{figure:naive_african_fitness_esc} shows a scatter-plot of the respective escape rate estimates. Figure \ref{figure:naive_african_fitness_rev} shows the corresponding plot for reversion rate estimates. Numbers after abbreviation of each epitope refers to the ordering in Tables \ref{SSITT_big_table} - \ref{bloemfontein_big_table}, as some epitopes have more than one escape mutation and/or associated restricting HLA. Spearman rank correlation coefficients with associated $p$-values are shown.}
\label{fig:app_fitness_africa}
\end{figure*}

\begin{figure*}
\centering
\includegraphics[width=0.75\textwidth]{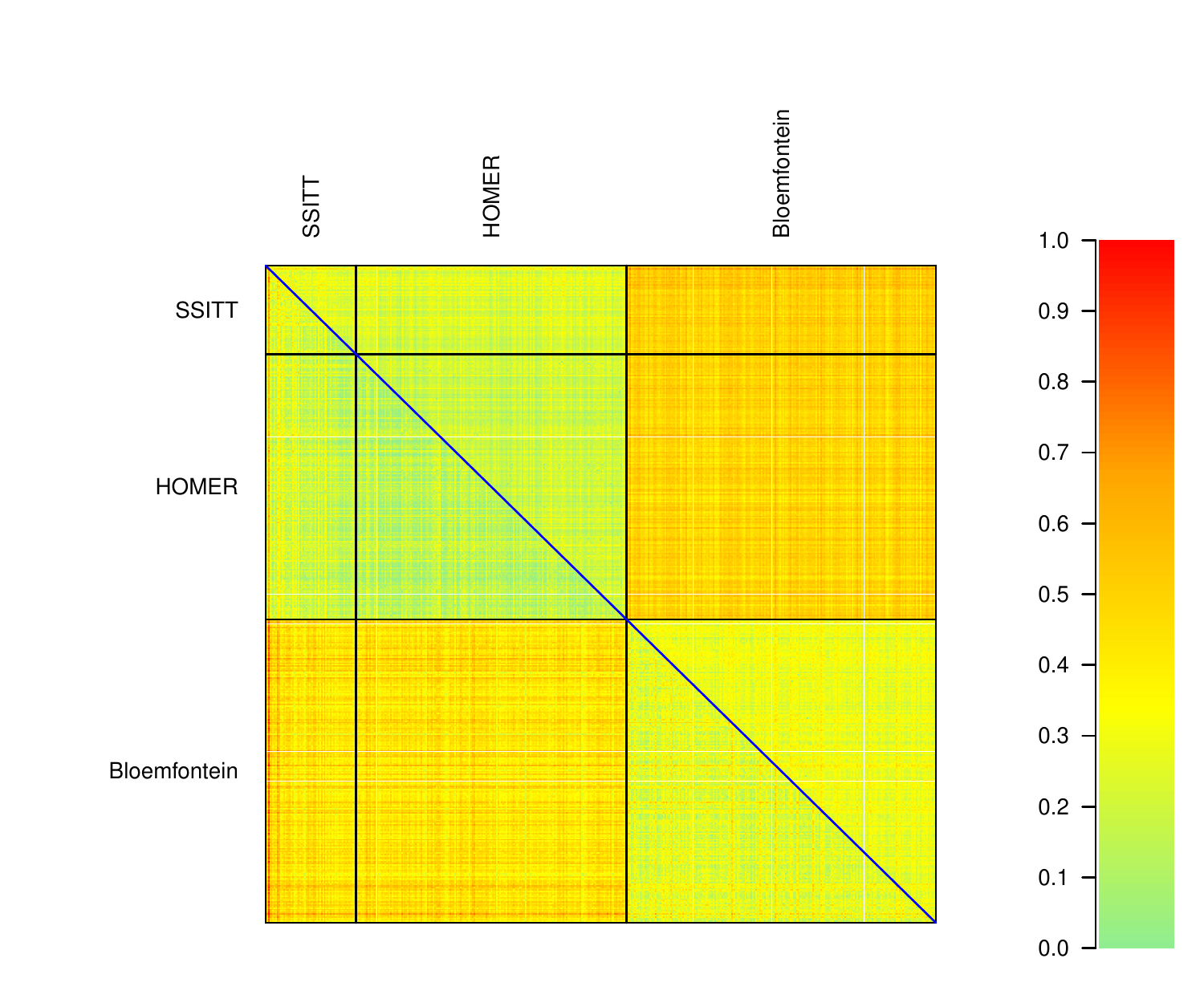}
\caption{{\bf Sequence divergence by Hamming distance}. We determine the Hamming distance between each pair of sequences in SSITT, HOMER and Bloemfontein for nucleotide and amino acid sequences. We then scale by sequence length to obtain a measure of sequence divergence. Nucleotide sequence divergence is shown above the leading diagonal, and amino acid sequence divergence is plotted below the leading diagonal. Black lines distinguish the SSITT, HOMER, Bloemfontein cohorts. The leading diagonal is shown in blue.}
\label{fig:sequence_divergence}
\end{figure*}

\begin{figure*}
\centering
\includegraphics[width=0.9\textwidth]{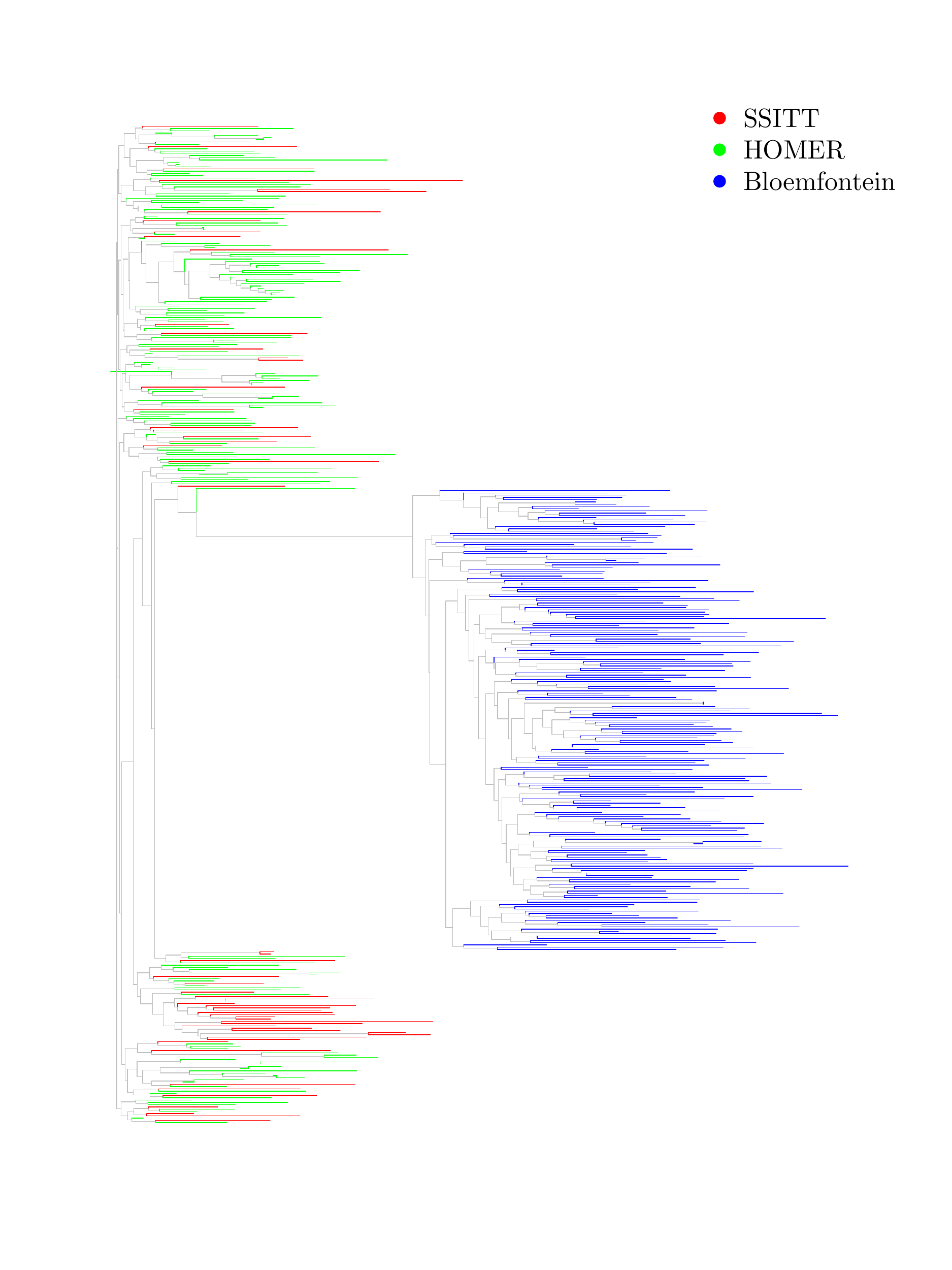}
\caption{{\bf Neighbour joining tree}. We show a neighbour joining tree obtained using the K81 \cite{citeulike:4210163} model applied to the largest region of {\it gag} over which $<10\%$ sequences contained a gap or an unknown nucleotide in the sequence alignment. Terminal branches are coloured by cohort according to the legend.}
\label{fig:neighbour_joining}
\end{figure*}
\end{document}